# Multi-objective Optimization of Savonius Wind Turbine

Seyed Ehsan Hosseini[a*], Omid Karimi[a], Mohammad Ali AsemanBakhsh[b]

[a] School of Mechanical Engineering, Iran University of Science and Technology, Tehran, Iran.

[b] Department of Mechanical Engineering, Shiraz Branch, Islamic Azad University, Shiraz, Iran.

*Corresponding author Email: ehsan_hosseini@alumni.iust.ac.ir

Tel.: +98-933-1887259

**Abstract:**

A numerical data set will be developed to assist in the design of Savonius wind turbines. The main objective of study is to improve Savonius turbine blade designs to increase torque coefficients, rotational speeds, and pressure coefficients. Simulating 3D models and validating them with wind tunnel data were part of the experimental design methodology. Multi-objective optimization is used to optimize turbine performance. Twist angle, aspect ratio, and overlap ratio are all important factors in determining the optimal torque and power coefficients. Data-driven objective functions were modeled using the group method of data handling (GMDH). Using an evolutionary Pareto-based optimization approach, polynomial models were used to plot Pareto fronts and TOPSIS to calculate optimal commercial points. The torque coefficient, rotational speed, and power coefficient are all improved by 13.74%, 0.071%, and 5.32%, respectively. As a result of the multi-objective optimization of the turbine, some significant characteristics of objective functions were discovered.

**Keywords:** Savonius wind turbine, modified NSGA-II, Artificial neural network, Multi-objective optimization.

1. **Introduction**

Due to environmental pollution, increased energy demand, and the depletion of fossil fuel resources, wind turbines have been gaining a great deal of attention. Since 2000, wind energy has contributed to an annual increase of 24% in electricity generation [1]. For the generation of electricity, there are two types of wind turbines: the horizontal axis wind turbine (HAWT) and the vertical axis wind turbine (VAWT) [2,3]. This article focuses on the VAWT, and in particular, the Savonius turbine. There are a variety of applications for the Savonius wind turbine, including water pumping, ventilation of buildings, electricity generation, and hybrid renewable energy systems (HRES) [4–8]. There have been numerous studies conducted on the geometric parameters of VAWTs [9–13]. An experimental study was conducted by Damak et al. [14] to optimize the rotor of a helical Savonius wind turbine. Based on their results, the modified twisted rotor displayed a higher power coefficient and static torque than the conventional twisted rotor. Montelpare et al. [15] studied and reported experimental data on a modified Savonius system generating street lighting at different twist angles. To increase the rotor's aerodynamic performance, a conveyor and a deflector are incorporated into their design. The results of Shaheen et al. [16] were compared with experimental data for the two-bucket Savonius wind turbine cluster in parallel and oblique positions. A numerical analysis of VAWT farms' performance parameters was also performed by Shaheen and Abdallah [17]. An experimental and numerical study was conducted by Lee et al. [18] to determine the performance and geometric specifications of a Savonius wind turbine under different twist angles. A constant torque coefficient was observed at twist angles greater than 90 degrees when numerical results obtained in an unsteady state were compared with experimental data. For the purpose of ventilating buildings, Tahani et al. [19] designed and constructed a Savonius VAWT with a direct discharge flow. Savonius wind turbines have been designed with a twist angle that reduces negative torque and improves performance. Kamoji et al. [20] examined the effects of overlap ratios, aspect ratios, and Reynolds numbers on the performance of the Savonius blade. It was found that the modified Savonius blade with an overlap ratio of 0 and an aspect ratio of 0.7 had the highest power coefficient of 0.21 at a given wind speed, achieving an overall power coefficient of 0.21. By comparing different blade shapes, Saha et al. [21] concluded that when two blades were arranged adjacently, the performance was better than one, two, or three blades. Additionally, the twisted blades performed better than the cylindrical blades.

In order to improve the performance of the Savonius turbine, Damak et al. [22] conducted an experimental study on the Savonius rotor with a 180° twist angle. A Savonius rotor with a twist angle of 180° outperformed a conventional Savonius rotor, according to their results. To improve the pressure coefficient of the Savonius turbine, Chan et al. [23] conducted studies on the optimization of the blade shape of the turbine. A new geometry was proposed and analyzed using genetic algorithm (GA) and computational fluid dynamics (CFD). Based on the results, the optimized blade performed better than the previous model. For the design of vertical-axis wind turbines, Bedon et al. [24] developed a database generation technique that was validated for symmetric profiles. There was an improvement in the performance of wind turbines as a result of the results. Jafaryar et al. [25] conducted a numerical study based on the central composite design (CCD) to obtain an optimal design for VAWT blades with an asymmetric geometry. According to their results, the maximum torque coefficient was obtained at a rotational speed of 450 rpm. Mohamad et al. [26] proposed the use of symmetric airfoil blades based on coupling optimization and CFD algorithms to increase the tangential force generated by the Wells monoplane turbine. Their optimization of the process resulted in an increase in the output power and efficiency of the machine.

According to the literature review, there are no studies that apply the TOPSIS and modified NSGA-II algorithms to the multi-objective optimization of Savonius wind turbines. The NSGA-II has been extensively studied as a multi-objective optimization algorithm [27–29]. Although this method has been presented previously based on GA, it has been widely used in the literature [30–33]. Savonius wind turbines have been studied experimentally and numerically, but little information is available regarding the role of geometrical and operational parameters. Specifically, based on the literature review presented above, there is a clear lack of research on multi-objective optimization of Savonius wind turbines using TOPSIS and modified NSGA-II which is fundamentally different from NSGA-II; this is a first-of-its-kind application of multi-objective optimization to a Savonius wind turbine.

An evaluation of the performance of a Savonius helical wind turbine is performed through experimental testing and numerical simulation in this study. Using available experimental data in the literature, the measured experimental data were validated. Afterward, the simulation results are validated against the experimental data. Based on the twist angle, aspect ratio, and overlap ratio as design variables, a series of simulations is conducted in order to obtain the polynomial of the models using GMDH. The objective functions selected are torque coefficient, rotational speed, and power coefficient. Using polynomial models, we then conducted a multi-objective optimization to optimize the turbine performance. TOPSIS is then used to determine the best trade-off points.

## 2. Experimental setup

### 2.1. Test sample and apparatus

Figure (1) shows a helical Savonius wind turbine that was tested in a wind tunnel at Radman San'at Co. This is an open-circuit suction tunnel with a test section area of 600x600 mm$^2$. For aerodynamic testing, it is capable of reaching a maximum speed of 60 m/s. Four different days were scheduled for the tests in order to ensure accuracy and repeatability. The rotational speed of the turbine was calculated using a tachometer. To detect the rotation of the shaft, axis, blade, or any other rotating object, the optical tachometer emits a red beam on a luminous label. RPM is then calculated and displayed as the rotational speed. The setup operates in the range of 2.5 to 99999 rpm. At speeds higher than 1000 rpm, the precision is 0.1 and 1 rpm, respectively. In front of the wind turbine, a standing hotwire was used to measure the airspeed. An S-shaped load cell was used to calculate weights up to 1000 N with a precision of 0.001.

### 2.2. Geometric model

Two twisted blades are attached to the central shaft for torque transmission. The turbine blades are made of PLA (Polylactic Acid) and have a diameter of 115 mm and a height of 162 mm. By keeping the extruded plastic warm, the 3D structure of the turbine shape can be manufactured using a 3D printer's robotic heat bed (Quantum Generous Pro).

The Quantum Generous Pro printer has an accuracy of 50 microns, and the largest dimensions are 300, 250, and 300 mm on the x, y, and z axes, respectively. As well, the 3D printing layer thickness has been set to 200 microns. The dimensions and operating parameters of the model, as well as the 3D wind turbine model, are used in the simulations presented in Table 1 and Fig. (1).

Initially, the wind speed was adjusted to 7 m/s by positioning the turbine in the middle of the test section. For accurate measurement of turbine performance at different torques, an optical tachometer was preferred over a dynamic tachometer. Dynamic calculations were used to calculate the turbine torque.

**Table 1** Geometric parameters of the helical rotor.

| Geometric parameters | Value |
|---|---|
| End plate ($D_0$) | 254 mm |
| Rotor diameter (D) | 230 mm |
| Rotor height (h) | 162 mm |
| Overlap ratio | 0 |
| Aspect ratio | 0.7 |
| Blade arc angle | 180° |
| Diameter of the blade | 115 mm |

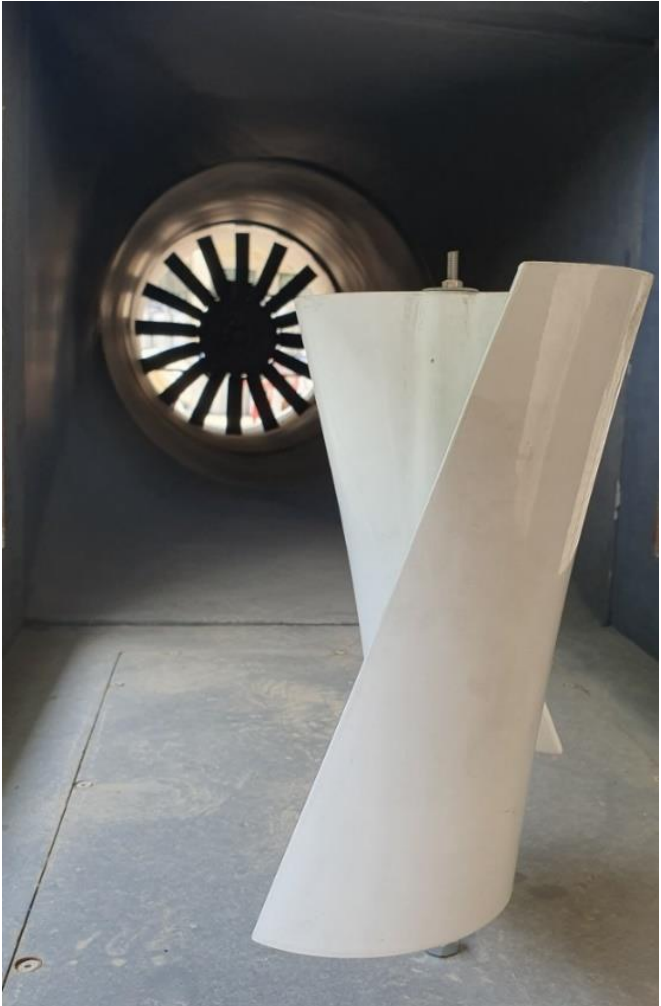 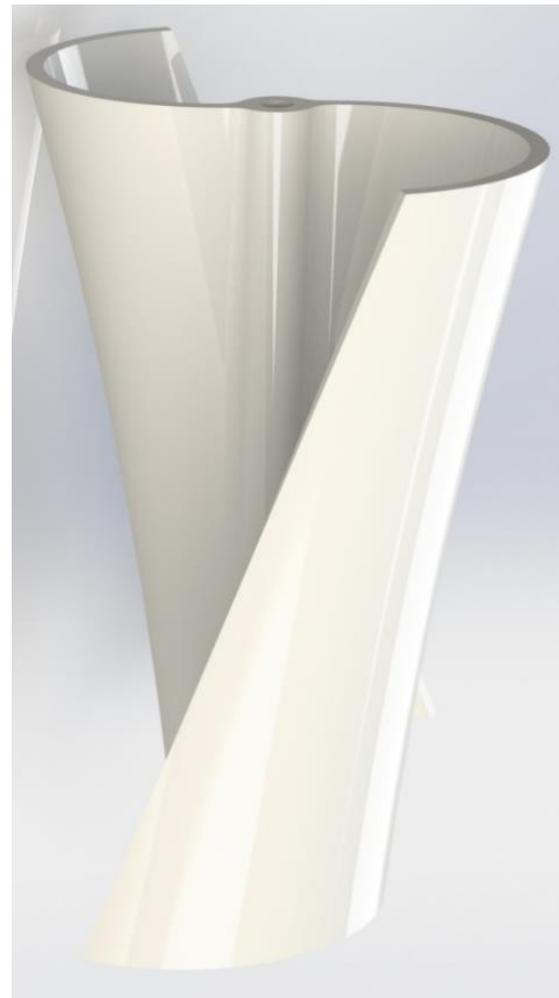

(a)          (b)

**Fig. 1** (a) Wind turbine blades in the test section, (b) blade geometry.

## 3. Numerical procedure

### 3.1. Governing equations

The governing equations for numerical simulation are based on the assumptions of incompressible fluids. The time-averaged Navier-Stokes equations used for determining the mass and momentum are as follows:

$$\frac{\partial \rho}{\partial t} + \nabla \cdot (\rho U) = 0 \tag{1}$$

$$\frac{\partial (\rho U)}{\partial t} + \nabla \cdot (\rho U \otimes U) = -\nabla p + \nabla \cdot \tau + S_m \tag{2}$$

where the stress tensor, τ, is related to the strain rate by:

$$\tau = \mu \left( \nabla U + (\nabla U)^T - \frac{2}{3} \delta \nabla \cdot U \right) \tag{3}$$

where $\mu$ is the fluid viscosity. The Boussinesq [34] concept which relates the Reynolds stresses to the mean velocity gradients with the turbulent eddy viscosity ($\mu_{turb}$) as the proportionality factor is used to model the stress tensor ($-\rho \overline{u'_i u'_j}$).

$$-\rho \overline{u'_i u'_j} = \left[ \mu_{turb} \left( \frac{\partial U_i}{\partial x_j} + \frac{\partial U_j}{\partial x_i} \right) \right] - \frac{2}{3} \rho k \delta_{ij} \tag{4}$$

where $k$ is the kinetic energy of turbulence, and all other symbols are given in the list of symbols in Table 2.

Finally, based on the available data in the literature [16,35], the shear stress transport (SST) model was selected as the most appropriate model. The transport equations for the kinetic energy of turbulence ($k$) and its turbulent frequency ($\omega$) are obtained from the following equation [36]:

$$k = \frac{1}{2} \sum_{i=1} \overline{u'^2_i} \tag{5}$$

$$\omega = \mu_{turb} \overline{\left(\frac{\partial u_i}{\partial x_j}\right)\left(\frac{\partial u_i}{\partial x_j}\right)} / \rho k \tag{6}$$

Where $u_i'$ is fluctuating velocity. The kinetic energy of turbulence deformation tensor is determined by:

$$\frac{\partial \rho k}{\partial t} + \frac{\partial}{\partial x_i}(\rho u_i k) = \frac{\partial}{\partial x_i}\left((\mu + \frac{\mu_{turb}}{\sigma_k})\frac{\partial k}{\partial x_i}\right) + P_k - \beta^* \rho \omega k \tag{7}$$

$P_k$ is production of turbulent kinetic energy due to interaction between relative flow and flow field:

$$P_k = \tau_{ij}\frac{\partial u_i}{\partial x_j} = \left[\mu_{turb}\left(\frac{\partial u_i}{\partial x_j} + \frac{\partial u_j}{\partial x_i} - \frac{2}{3}\frac{\partial u_k}{\partial x_k}\delta_{ij}\right) - \frac{2}{3}\rho k \delta_{ij}\right]\frac{\partial u_i}{\partial x_j} \tag{8}$$

In which $\beta^* \rho \omega k$ is turbulent kinetic depreciation. The turbulent frequency deformation tensor is given by:

$$\frac{\partial \rho \omega}{\partial t} + \frac{\partial}{\partial x_i}(\rho u_i \omega) = \frac{\partial}{\partial x_i}\left\{(\mu + \frac{\mu_{turb}}{\sigma_\omega})\frac{\partial \omega}{\partial x_i}\right\} + \frac{\rho \gamma}{\mu_{turb}}\tau_{ij}\frac{\partial u_i}{\partial x_j} - \rho \beta \omega^2 +$$
$$2\rho(1-F_1)\sigma_{w2}\frac{1}{\omega}\frac{\partial k}{\partial x_j}\frac{\partial \omega}{\partial x_j} \tag{9}$$

where $\phi = F_1\phi_1 + (1-F_1)\phi_2$ is constant. Model constants are listed in Table 3. $F_1$ is a blending function composition which is given by:

$$F_1 = \tanh(\arg_1^4) \tag{10}$$

Also:

$$\arg_1 = \min\left[\max\left(\frac{\sqrt{k}}{0.09\omega y}, \frac{500\mu}{y^2 \omega \rho}\right), \frac{4\rho \sigma_{w2} k}{CD_{k\omega} y^2}\right] \tag{11}$$

where $y$ is a distance to the nearest wall, $CD_{k\omega}$ is the positive portion of the cross-diffusion term of Eq. 12:

$$CD_{k\omega} = \max(2\rho\sigma_{w2}\frac{1}{\omega}\frac{\partial k}{\partial x_j}\frac{\partial \omega}{\partial x_j}, 10^{-20}) \qquad (12)$$

The turbulent viscosity is obtained using a limiter. Menter [37] has argued that the use of this limiter does not increase the turbulent viscosity in regions close to the stagnation point:

$$\mu_{turb} = \frac{0.31\rho k}{\max(0.31\omega, \Omega F_2)} \qquad (13)$$

where $\Omega$ is the absolute value of vorticity given by $\Omega = \sqrt{2W_{ij}W_{ij}}$ where $W_{ij} = \frac{1}{2}\left(\frac{\partial u_i}{\partial x_j} - \frac{\partial u_j}{\partial x_i}\right)$.

The blending function $F_2$ can be expressed by:

$$F_2 = \tanh(\arg_2^2)$$
$$\arg_2 = \max(\frac{2\sqrt{k}}{0.09\omega y}, \frac{500\mu}{y^2\omega\rho}) \qquad (14)$$

F1 and F2 are the blending functions which are based on distance from nearest wall to blend the near-wall k-ω model with the away from-wall k-ε closure \ (recast into k-ω variables), this being a fundamental attribute of the SST model. In the present work, these two functions redefined in a form that replaces wall distance with a local representation.

**Table 2** Description of symbols.

| | | | |
|---|---|---|---|
| $k$ | Turbulent kinetic energy (m$^2$/s$^2$) | $F_1$ | Blending function composition |
| $\mu$ | Fluid viscosity (Pa s) | $y$ | Distance to the nearest wall (m) |
| $P$ | Pressure (Pa) | $\tau$ | Stress tensor |
| $u'$ | Fluctuating velocity (m/s) | $\mu_{turb}$ | Turbulent viscosity (Pa s) |
| $\rho$ | Density (kg/m$^3$) | $P_k$ | Rate of shear production of $k$ |
| $U$ | Flow velocity (m/s) | $\delta$ | Kronecker delta |
| $t$ | Time | $\omega$ | Turbulent frequency (s$^{-1}$) |
| $\Omega$ | absolute value of vorticity | | |

**Table 3** Constants of set 1 ($\phi_1$) and set 2 ($\phi_2$).

| $\sigma_{k1}$ | $\sigma_{\omega1}$ | $\beta_1$ | $\beta^*$ | $\gamma_1$ |
|---|---|---|---|---|
| 0.85 | 0.5 | 0.075 | 0.09 | 5/9 |
| $\sigma_{k2}$ | $\sigma_{\omega2}$ | $\beta_2$ | $\beta^*$ | $\gamma_2$ |
| 1.0 | 0.856 | 0.0828 | 0.09 | 0.44 |

## 3.2. Boundary conditions

The turbine performance should be simulated in a rectangular duct with a cross-section of (3 × 3.5) m and a length of 7 m. The rotor interface consists of a rotational domain and a stationary zone. Wind speed is specified at the inlet, pressure is specified at the outlet, and all other boundaries are specified as non-slip.

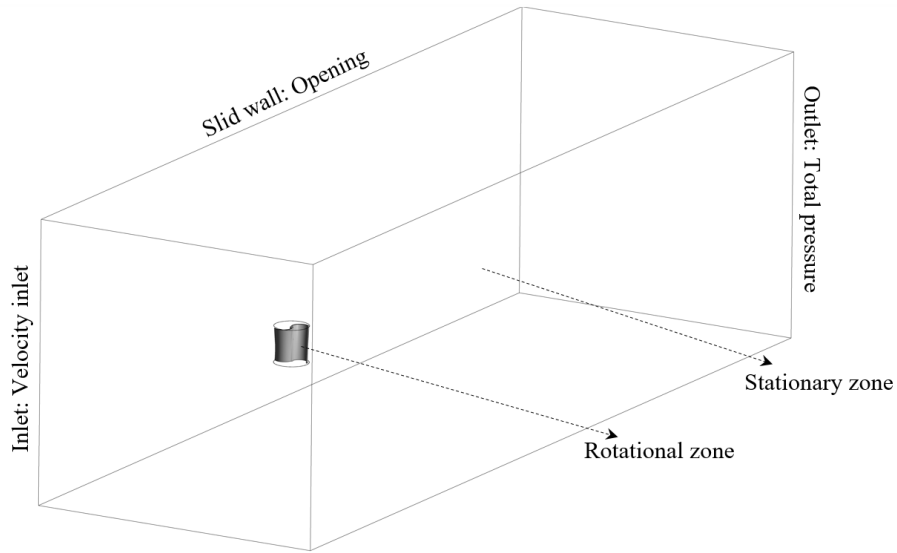

**Fig. 2** Computational domain and framework.

### 3.3. Solver settings

Based on ANSYS-CFX (Version 16.1), the governing equations were numerically simulated using a finite-volume method. A high-resolution scheme, SIMPLE scheme (Semi-Implicit Method for Pressure Linked Equations), was used to discretize the advection term. Furthermore, a convergence criterion of $10^{-6}$ was used in this study.

### 2.4. Turbine parameters

The following equations were used for the calculation of the power and torque coefficients, denoted by $C_p$ and $C_T$, respectively. The power coefficient is defined as the ratio of the power generated by the wind turbine ($P_w$) to the power available in the wind ($P_a$):

$$C_p = \frac{P_w}{P_a} = \frac{\omega T}{\frac{1}{2}\rho h D U^3} \tag{15}$$

where $\omega$ represents the rotational speed of the turbine, $T$ the power generated by the turbine, $\rho$ the air density, $U$ the velocity of the free flow colliding the turbine blades, and $h$ and $D$ respectively show the height and diameter of the turbine. Similarly, the torque coefficient is defined as the ratio of the generated torque ($T$) to the theoretical torque ($T_w$):

$$C_T = \frac{T}{T_w} = \frac{T}{\frac{1}{2}\rho A_s d U^2} \tag{16}$$

where $A_s$ is the vertically projected area of the blade, calculated by $h \times D$, where $d$ is the diameter of each blade.

The tip speed ratio (α) is also defined as:

$$\lambda = \frac{v_{rotor}}{u_\infty} = \frac{|\omega| d}{u_\infty} \tag{17}$$

The speed of the blade at its tip can be calculated given the rotational speed and diameter of each blade. This speed is proportional to the operating point of the wind turbine for the maximum generated power. The aspect ratio, denoted by $α=h/D$, is defined as the ratio of the height of the turbine blades to the turbine diameter.

As shown in Fig. (3) the case where the blades diverge along the direction perpendicular to the curvature line at the connection point, the overlap ratio ($δ=e/d$) is defined as the ratio of the spacing of the blades along this direction ($e$) to the length of the blade cord.

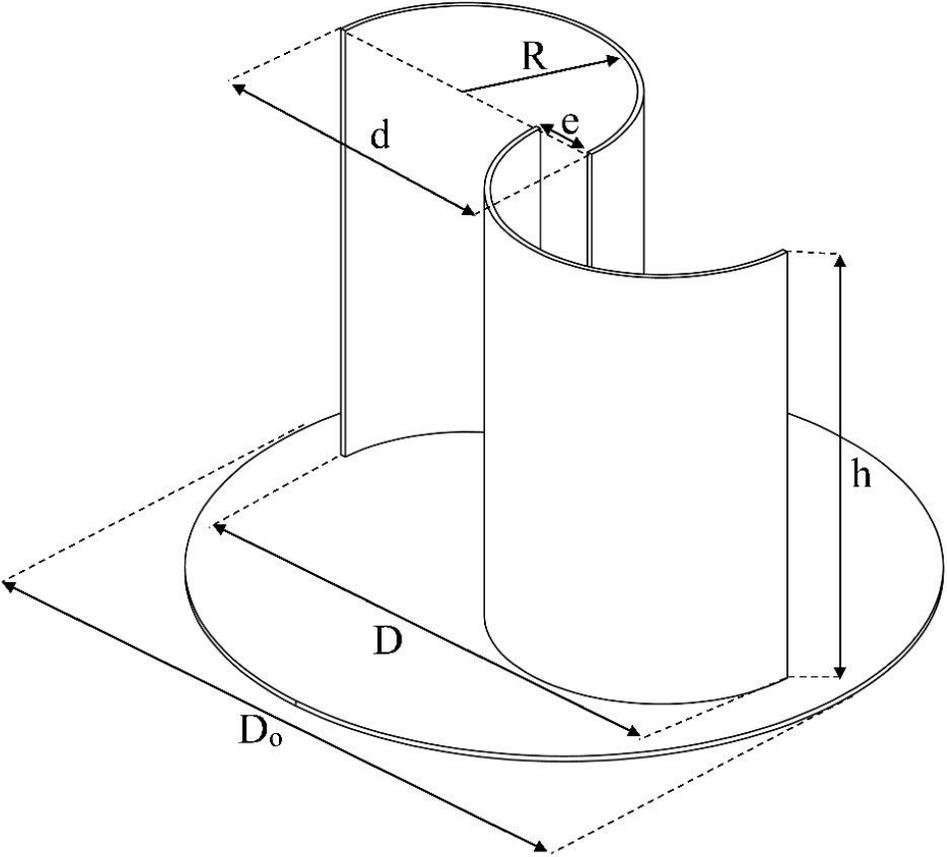

**Fig. 3** Scheme of variables.

## 2.5. Grid generation and grid independence check

Due to the complexity of the geometric shapes investigated in this study, ANSYS ICEM (Version 16.1) was used to generate unstructured tetrahedral and prismatic elements for modeling the Savonius wind turbine. The small gap between the blade and axis zones also required mesh refinements to capture the flow details more accurately. Fig. (4) illustrates the schematic of the mesh used in this study. It is necessary to solve the governing equations for different numbers of grid points to ensure that the results are mesh-independent, and Table 4 presents the results of numerical simulation for wind turbine pressure coefficient, torque coefficient, rotational speed, and CPU times. Note that the results were obtained using a Core i7 (8 cores) CPU running at 2.6 GHz. According to the table, no changes can be observed in these parameters when considering 8,510325 cells. For numerical simulations, the grid with this number of cells is used.

**Table 4** Grid independency study.

| Number of cells | $C_p$ | $C_T$ | $\omega$ | CPU times |
|---|---|---|---|---|
| 3569842 | 0.17325 | 0.14659 | 39.5621 | 7 h, 36 min, 53 sec |
| 5649551 | 0.16491 | 0.15682 | 36.6586 | 9 h, 15 min, 12 sec |
| 6589425 | 0.13658 | 0.15985 | 35.8954 | 9 h, 57 min, 30 sec |
| 8510325 | 0.12735 | 0.174851 | 35.6972 | 10 h, 49 min, 33 sec |
| 9786521 | 0.12732 | 0.174849 | 35.6986 | 11h, 22 min, 10 sec |

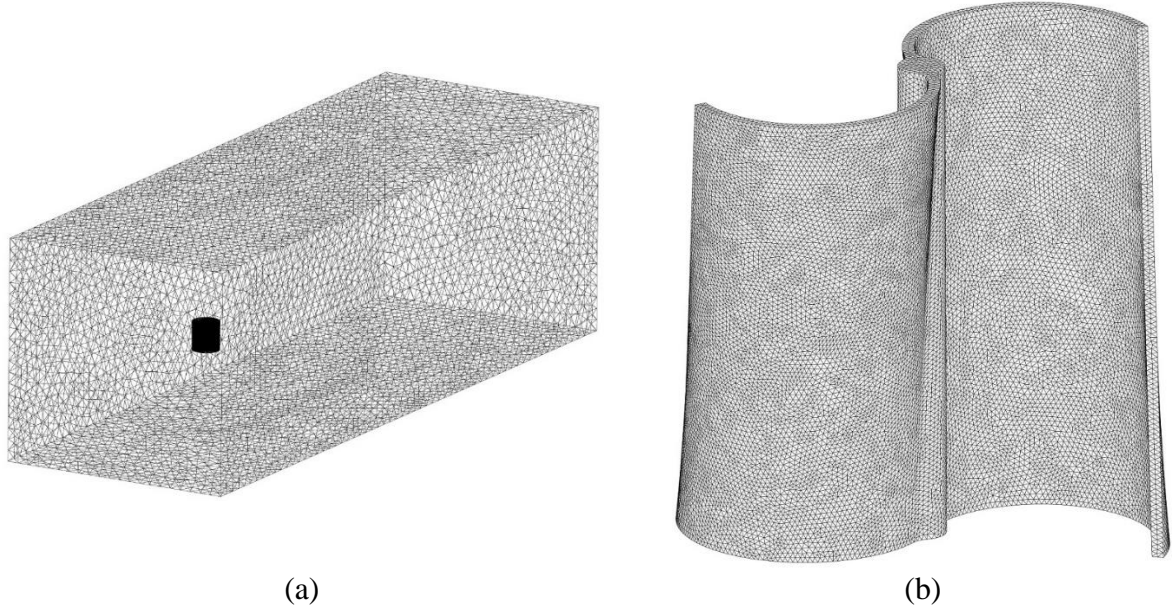

(a)                                               (b)
**Fig. 4** (a) Schematic of the unstructured mesh of the whole computational domain, (b) the blade unstructured mesh.

## 4. Experimental data and numerical results validation

It was first necessary to examine the accuracy of the measurement system to evaluate the power and torque coefficients of the turbine as well as the rotational speed. Using a 3D printer, the blade was fabricated according to the specifications in Table 1. Fig. (5) compares the power and torque coefficients with those reported in [14]. The Savonius wind turbine performance is characterized by the pressure coefficient ($\varphi$) and the torque coefficient ($\psi$) given by Eqs. (15) and (16). In view of the strong dependence of the Savonius wind turbine performance on wind speed, the effect of wind speed is examined for wind speeds of 4.9, 7, 8, and 8.8 m/s. The results of this study differ slightly from those of the reference study [14] due to friction, different power and torque measurement systems, and different test section specifications. As part of these differences, oscillations are included along with the diagram as well as within the TSR interval of the turbine performance. Therefore, the measurement error of power and torque coefficients at the tip speed ratios between 0.7 and 0.9 is less significant than at the other tip speed ratios, possibly because of bearing friction embedded beneath the turbine. Further, when the speed ratio is low, the effect of wind speed on the pressure coefficient is less pronounced than when the speed ratio is high. By reducing the tip speed ratio, it is evident that the torque coefficient increases. Specifically, the torque coefficient occurs at C = 0.15 and 0.35. The maximum error percentage for the power and torque coefficients is 3% and 4%, respectively, for tip speed ratio values of 0.8 and 0.61, for a wind speed of 7 m/s.

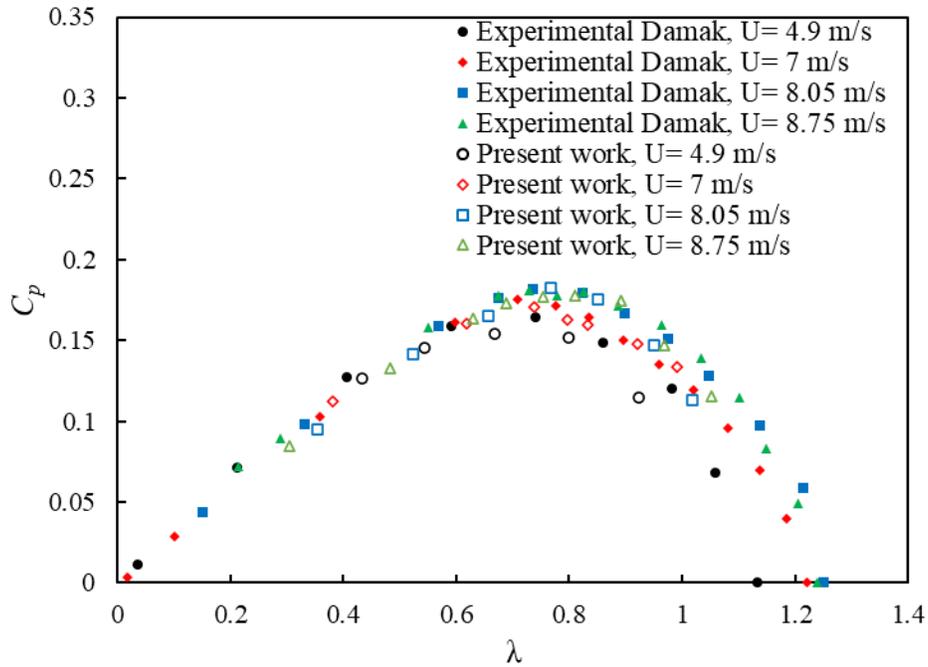

(a)

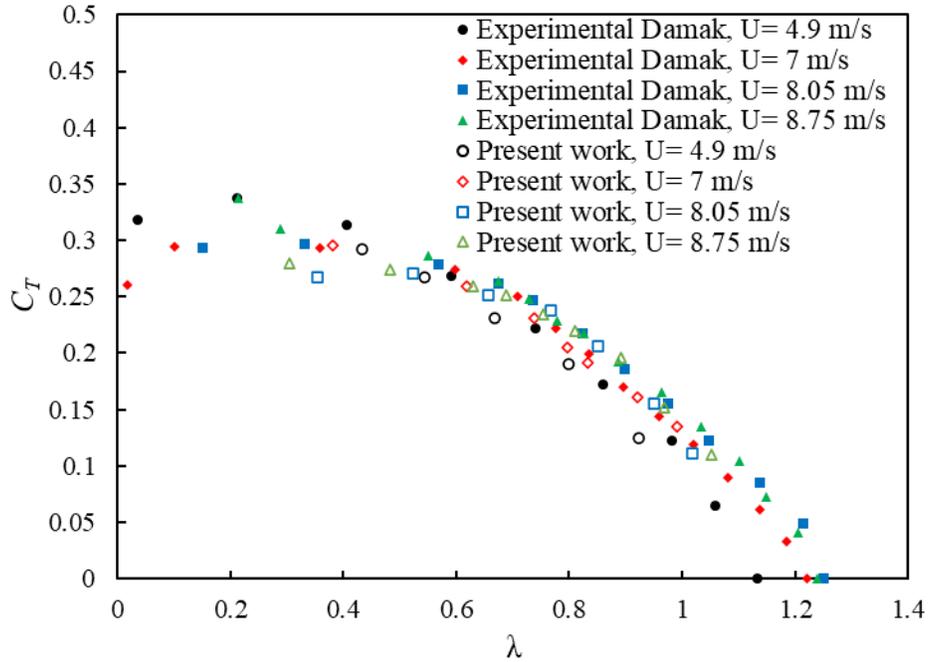

(b)

**Fig. 5** (a) Variation of power coefficient $C_p$, and (b) variation of torque coefficient $C_T$ due to tip speed ratio in validation.

Figure (6) presents a comparison between the numerical and experimental results. As shown in Fig. (6a), as the load increases, the power coefficient reaches its maximum value of 0.16 at a tip speed ratio of 0.74, exhibiting a linear behavior up to the tip speed ratio of 0.47. The power coefficient significantly decreases when the turbine is excessively loaded and when friction is excessive in the transmission system. Numerical results are in very good agreement with experimental data, which demonstrates the accuracy of the numerical method and assumptions used. It is also observed that the difference between the experimental and numerical results is the highest, with an error percentage of approximately 6%. The torque coefficient of the turbine is shown in Fig. (6b). The torque coefficient increases up to 0.27 with increasing applied load up to a tip speed ratio of 0.4 and then decreases slightly after that. According to the aforementioned relationships, Fig. (6c) illustrates a linear relationship between the tip speed ratio and the rotational speed. Maximum and average relative errors are 4.68% and 2.85%, respectively.

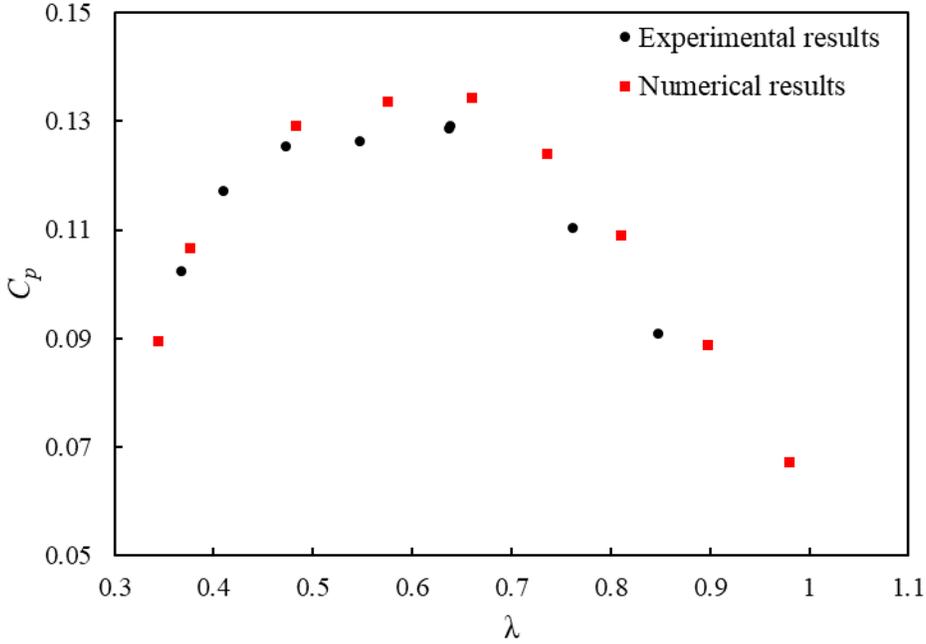

(a)

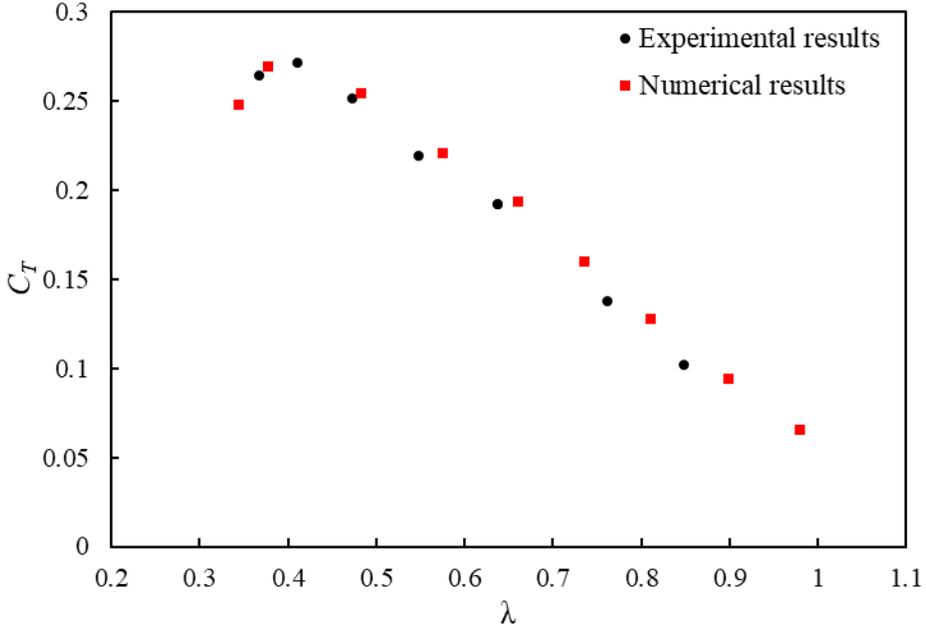

(b)

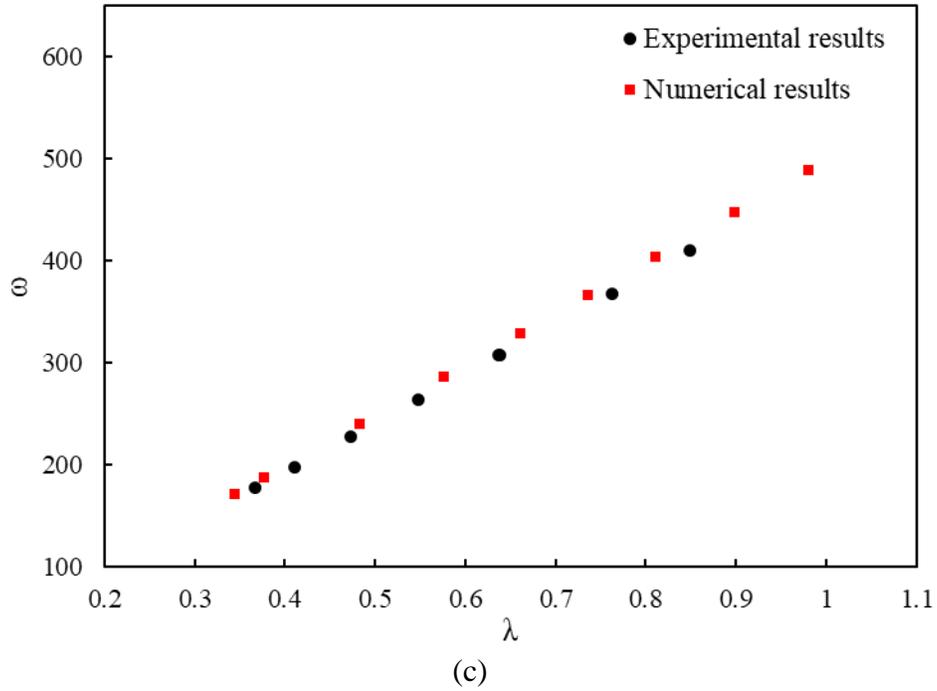

(c)

**Fig. 6** (a) Variation of power coefficient $C_p$, (b) torque coefficient $C_T$, and (c) rotational speed ω versus tip speed ratio λ.

## 5. Modeling using GMDH

In simplest terms, artificial neural networks (ANNs) are new computational systems and methods for machine learning and knowledge representation that aim at applying the acquired knowledge for predicting the output of complex systems. These networks are partly based on the performance of biological neural networks when it comes to processing data and information for learning and knowledge creation. This idea focuses on creating new structures for the information processing system. Through electromagnetic communication, neurons, which are highly interconnected processing elements, solve problems in a coordinated manner. Table 5 lists the main design parameters for artificial neural networks. The turbine construction takes into account all operating conditions of the design points. This resulted in 60 different geometries being created using the specifications reported in Table 6 (see Appendix 1). From our CFD analysis, we have obtained these different geometries separately. Blade twist angle, aspect ratio, and overlap ratio are three of the most important design parameters affecting turbine performance. A modified NSGA-II algorithm will be used for Pareto-based multi-objective optimization of the turbine.

**Table 5** Main design variables and their ranges in performance.

| Design variables | Lower bounds | Upper bounds |
|---|---|---|
| $\varphi$ | 0 | 60 |
| $\alpha$ | 0.8 | 1.2 |
| $\delta$ | 0 | 0.24 |

Ivankhenko [38] developed the GMDH method to model complex systems involving multiple inputs and a single output. As a matter of fact, this is a mathematical polynomial algorithm capable of designing a data structure for relating objective functions to design parameters. One of the best and most accurate methods for solving such data is the GMDH algorithm. The network describes the approximate output function from inputs (X = (x1, x2, x3, ..., xn)) with the most minor error compared to the actual output by combining second-order polynomials derived from all neurons. As a result, Equation (18) can be used to express actual results for $M$ experimental data including n inputs and a single output. For each input vector x, the output is expressed as follows:

$$y_i = f(x_{i1}, x_{i1}, x_{i1}, \ldots, x_{in}) \qquad i = 1, 2, \ldots, n \tag{18}$$

$$\hat{y}_i = \hat{f}(x_{i1}, x_{i1}, x_{i1}, \ldots, x_{in}) \qquad i = 1, 2, \ldots, n \tag{19}$$

In fact, the squared error between the real and predicted values should be minimized:

$$\sum_{i=1}^{M}(\hat{y}_i - y_i)^2 \to min \tag{20}$$

The polynomial function relating the design variables to the objective functions is obtained from Equation (21):

$$y = a_0 + \sum_{i=1}^{n} a_i x_i + \sum_{i=1}^{n}\sum_{j=1}^{n} a_{ij} x_i x_j + \sum_{i=1}^{n}\sum_{j=1}^{n}\sum_{k=1}^{n} a_{ijk} x_i x_j x_k + \ldots \tag{21}$$

This polynomial is the most common bivariate second-order polynomial:

$$\hat{y} = G(x_i, x_j) = a_0 + a_1 x_i + a_2 x_j + a_3 x_i^2 + a_4 x_j^2 + a_5 x_i x_j \tag{22}$$

These unknown coefficients are obtained using regression techniques by minimizing the difference between real and calculated outputs.

Moreover, the number of neurons in the second layer equals $C_2^n$ ($\binom{2}{n} = \dfrac{n(n-1)}{2}$). Equation (23) is used to obtain the unknown coefficients:

$$\{(y_i, x_{ip}, x_{iq}) \mid (i = 1, 2, \ldots, M)\} \tag{23}$$

The matrix form of these equations is presented in Equation (24):

$$Aa = Y$$
$$a = \{a_0, a_1, \ldots, a_5\} \tag{24}$$
$$Y = \{y_1, y_2, y_3, \ldots, y_M\}^{\mathrm{T}}$$

The multiple regression analysis is used in this method, to obtain the matrix coefficients in the form of Eq. 25:

$$A = \begin{bmatrix} 1 & x_{1p} & x_{1q} & x_{1p}^2 & x_{1q}^2 & x_{1p} x_{1q} \\ 1 & x_{2p} & x_{2q} & x_{2p}^2 & x_{2q}^2 & x_{2p} x_{2q} \\ \vdots & \vdots & \vdots & \vdots & \vdots & \vdots \\ 1 & x_{Mp} & x_{Mq} & x_{Mp}^2 & x_{Mq}^2 & x_{2p} x_{2q} \end{bmatrix} \tag{25}$$

To use the neural network, the data is divided into three categories. There are three categories of input parameters in the first category: twist angle, aspect ratio, and overlap ratio, and torque coefficient as the objective function. In the second and third categories, the input data are identical to the first category, but the rotational speed and power coefficient are considered the objective functions. The power, torque, and rotational speed of the wind turbine should be maximized to increase the efficiency of the wind turbine and optimize its performance. It is important to note that the ANN minimizes the functions. As a result, the torque and power coefficients, as well as the rotational speed, are inverted before being applied to the neural network. To facilitate calculations and improve convergence, 1/pressure coefficient, 1/torque coefficient, and 1/rotational speed are modeled as outputs using the GMDH algorithm.

Figure (7) displays the structure of the ANN for the objective functions, namely the power and torque coefficients and the rotational speed. The structure is the same for all three cases.

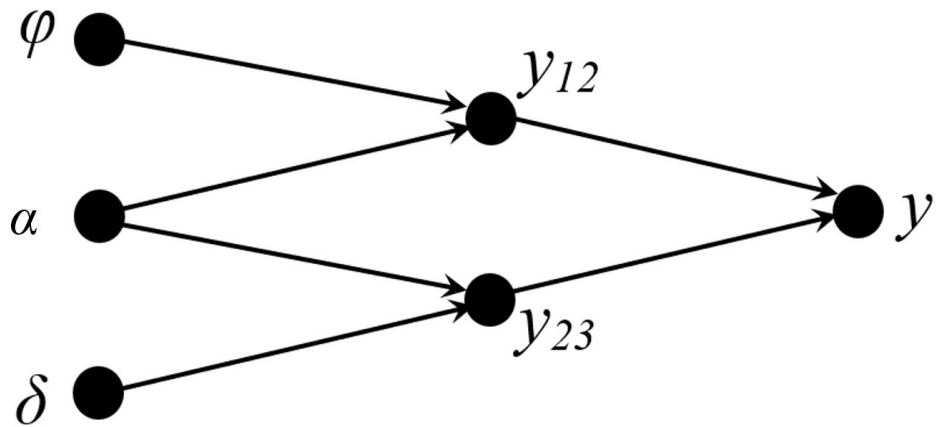

**Fig. 7** The artificial neural network structure for objective functions.

The ANN polynomials for the power and torque coefficients and the rotational speed of the turbine are as Equations (27- 29).

$$\frac{1}{C_p} = 54.618 - 2.648Y_{12} - 8.504Y_{23} + 0.0204Y_{12}^2 + 0.496Y_{23}^2 - 0.027Y_{12}Y_{23}$$
$$Y_{12} = 26.835 + 0.018\varphi - 39.259\alpha + 0.001\varphi^2 + 21.933\alpha^2 - 0.098\varphi\alpha \qquad (27)$$
$$Y_{23} = 28.352 - 40.508\alpha + 2.976\delta + 20.701\alpha^2 + 19.044\delta^2 - 4.633\alpha\delta$$

$$\frac{1}{C_T} = -12.181 - 0.006Y_{12} + 4.435Y_{23} - 0.157Y_{12}^2 - 0.534Y_{23}^2 + 0.456Y_{12}Y_{23}$$
$$Y_{12} = 25.664 - 0.017\varphi - 41.463\alpha + 0.001\varphi^2 + 22.735\alpha^2 - 0.069\varphi\alpha \qquad (28)$$
$$Y_{23} = 25.709 - 40.395\alpha - 9.284\delta + 20.834\alpha^2 + 41.163\delta^2 - 0.445\alpha\delta$$

$$\frac{1}{\omega} = 0.305 - 3.828Y_{12} - 14.972Y_{23} + 0.168Y_{12}^2 + 178.283Y_{23}^2 + 140.395Y_{12}Y_{23}$$
$$Y_{12} = -0.00003 + 0.0002\varphi + 0.058\alpha - 0.0000008\varphi^2 - 0.0293\alpha^2 - 0.0001\varphi\alpha \qquad (29)$$
$$Y_{23} = 0.023 + 0.012\alpha - 0.018\delta - 0.006\alpha^2 + 0.231\delta^2 - 0.005\alpha\delta$$

According to Fig. (8), the power and torque coefficients and rotational speed predicted by the GMDH are compared to those predicted by the CFD. The results of the ANN are consistent with those of the CFD. Using the evolved group method of data handling type neural networks to obtain polynomial equations for ANNs, we applied multi-objective optimization to simultaneously vary blade twist angle, aspect ratio, and overlap ratio functions. Fig. (8) clearly demonstrates that the polynomial equations predict the targets for the testing data that have not been utilized during training. These evolved GMDH-type neural networks are tested on two separate sets of data, one for training and one for testing, in order to demonstrate their prediction ability. In order to train the neural network models, 42 out of 60 input-output data sets are used. In the training process, the test set consists of 18 unpredictable input-output data samples, which are used only for testing the predictive ability of the evolved GMDH-type neural network models.

The performance of the ANN is evaluated using the absolute fraction of variance ($R^2$), root-mean-square error (*RMSE*), and mean relative error (*MRE*), which are defined as follows:

$$R^2 = 1 - \frac{\sum_j (t_j - o_j)^2}{\sum_j o_j^2}$$

$$RMSE = \sqrt{\frac{\sum_j (t_j - o_j)^2}{P}} \tag{30}$$

$$MRE\,(\%) = \frac{1}{P} \sum_j \left| \frac{(t_j - o_j)}{o_j} \times 100 \right|$$

where $t$ is the target value, $o$ is the output value and $P$ is the pattern number. The results show that the model confirms the observations (Table 7).

The convergence history of RMSE for the power and torque coefficients and the rotational speed predicted by the GMDH are shown in Fig. (9). It is worth mentioning that the minimum of the bias for the power and torque coefficients and the rotational speed are 0.08, 0.002, and 0.056, respectively.

**Table 7** The values of the $R^2$, *RMSE*, and *MRE* for evaluating the artificial neural network performance.

| Output | $R^2$ | RMSE | MRE |
|---|---|---|---|
| $1/C_p$ | 0.9812 | 1.0092 | 12.2264 |
| $1/C_T$ | 0.9862 | 1.1376 | 5.40192 |
| $1/\omega$ | 0.9967 | 0.1719 | 9.35679 |

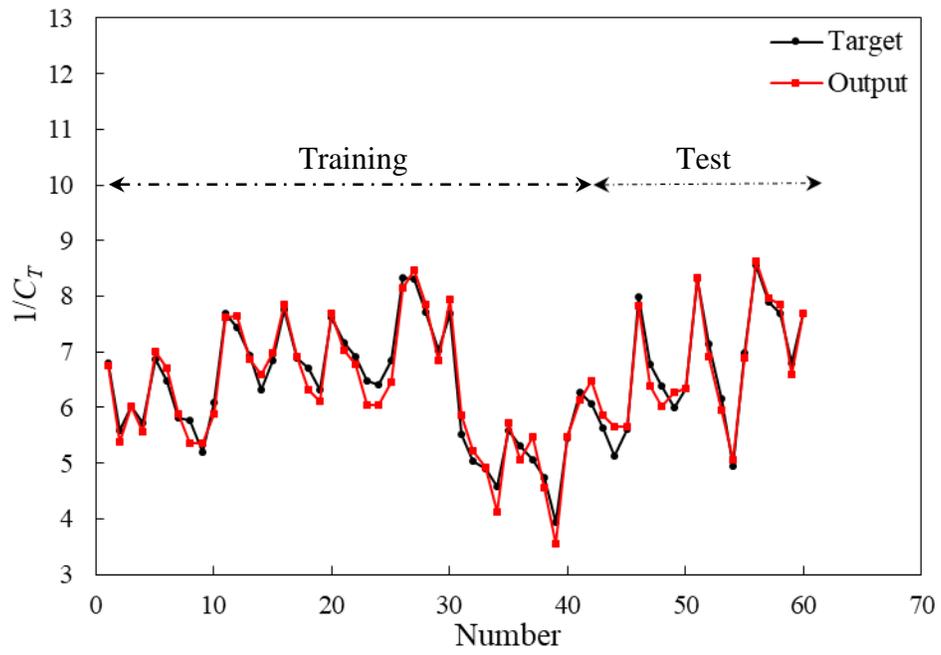

(a)

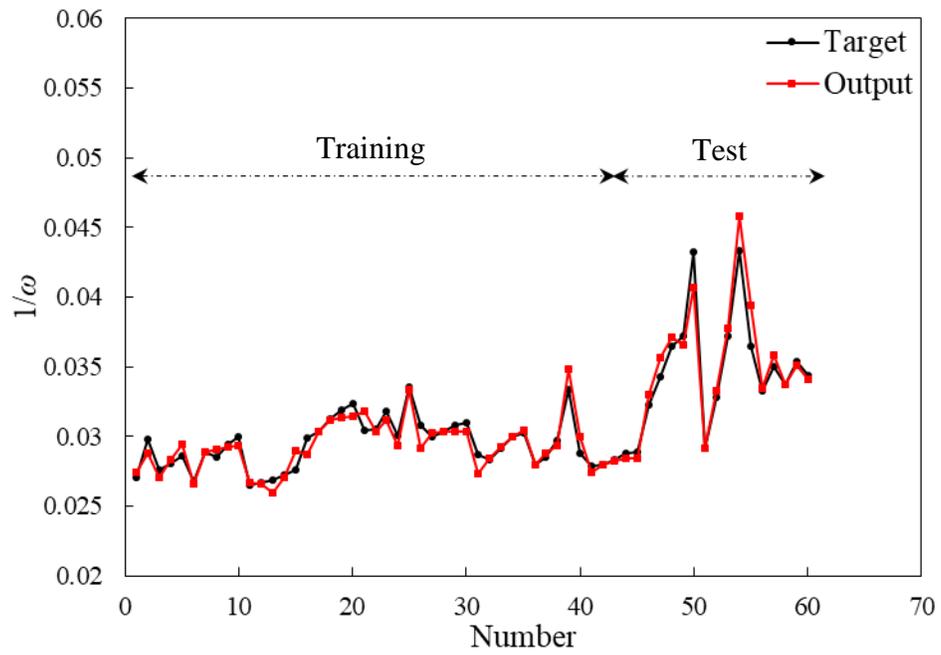

(b)

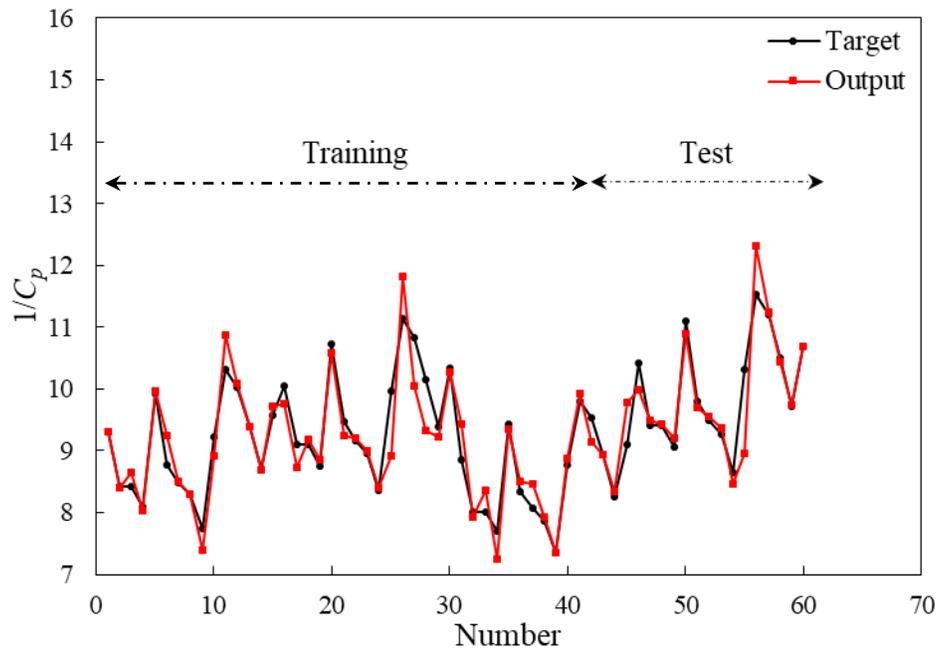

(c)

**Fig. 8** (a) Comparison of the predicted results by artificial neural network with training and test data sets for $1/C_T$, (b) $1/\omega$, and (c) $1/C_p$

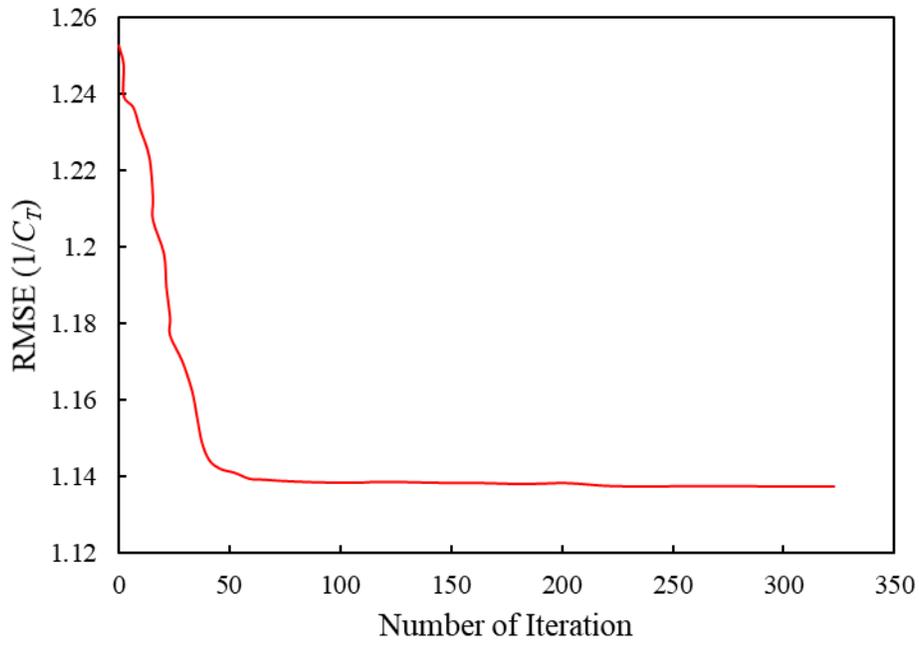

(a)

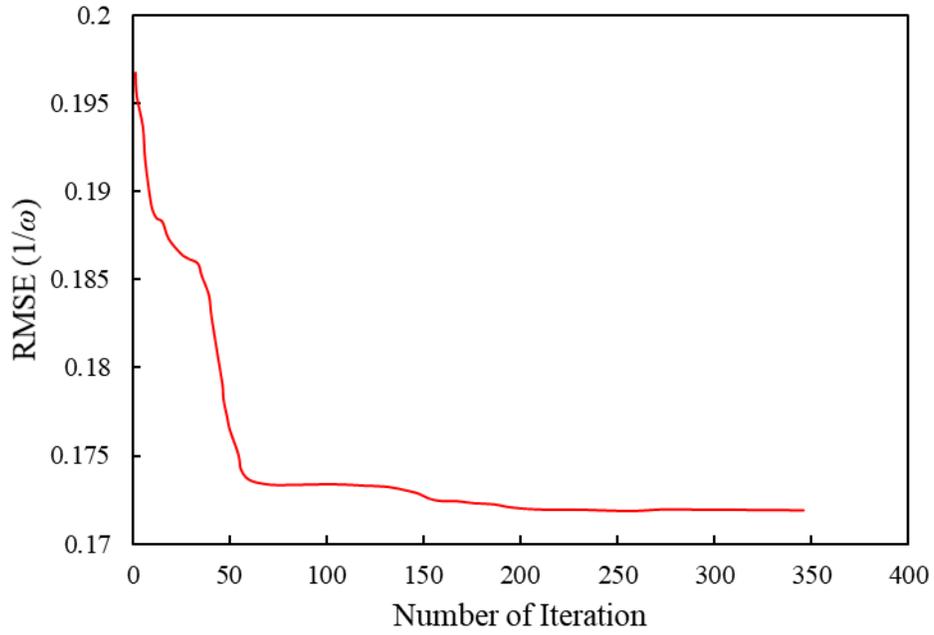

(b)

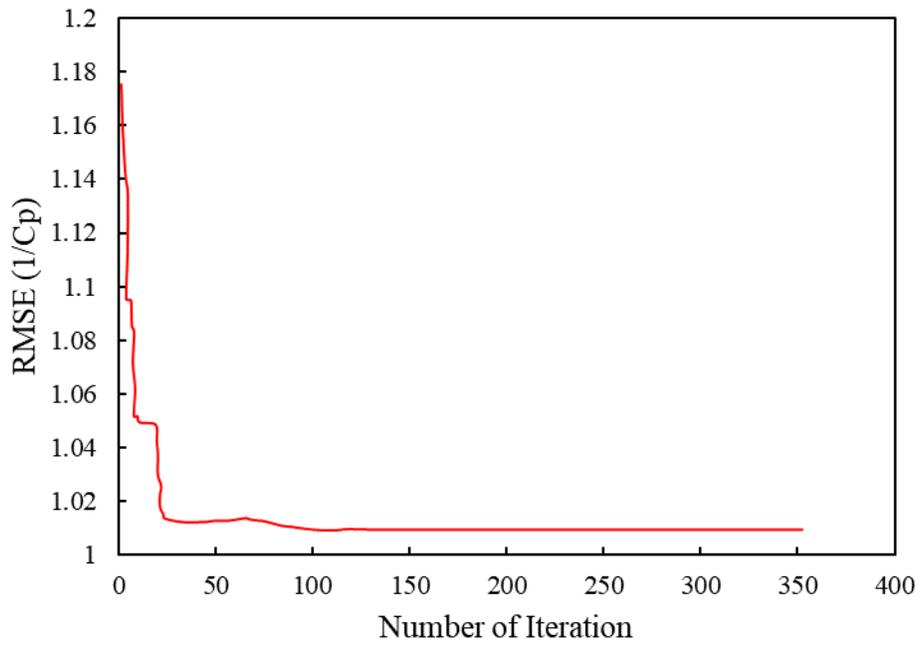

(c)

**Fig. 9** (a) Convergence history of RMSE for $1/C_T$, (b) $1/\omega$, and (c) $1/C_p$

## 6. Multi-objective optimization

The multi-objective evolutionary optimization algorithms are becoming an important modeling tool for solving multi-objective problems, in addition to their ever-increasing applications in different sciences. An improved VAWT was modeled using the modified multi-objective NSGA-II (Fig. (1b)). A multi-objective optimization algorithm that is widely used is the modified NSGA-II. Furthermore, it is considered to be one of the most fundamental algorithms in the field of multi-objective evolutionary optimization. The objective of this optimization is to obtain the best combination of design parameters.

According to the results and the relations used for wind turbine performance, the torque coefficient, rotational speed, and consequently power coefficient are affected by the twist angle, aspect ratio, and overlap ratio.

A multi-objective optimization algorithm can be mathematically defined as the process of searching for the vector of design variables ($X^* = [x_1^*, x_2^*, ..., x_n^*]$, $X^* \in \Re^n$), such that the limitations imposed by *m* inequalities and *p* equalities are met:

$$g_i(X) \leq 0, \quad i = 1 \text{ to } m \tag{31}$$
$$h_j(X) = 0, \quad j = 1 \text{ to } p \tag{32}$$

To optimize the vector of objective functions:
$$F(X) = [f_1(X), f_2(X), ..., f_k(X)]^T, \quad F(X) \in \Re^n \tag{33}$$

To put it another way, all objective functions in the problem should be optimized simultaneously. Due to the conflict between these objective functions, an increase in one will affect the others. There is no single optimal solution with respect to all objective functions, but rather a set of solutions known as the Pareto front [39].

Despite the fact that these optimal solutions do not dominate one another, they are superior to other solutions in the domain of objective functions. Choosing a better value for one objective function can result in a worse value for another objective function. Consequently, all objectives cannot be improved by changing the value of the design variables for these objective functions at the same time. According to the Pareto front, each solution has an objective function lower than other.

For the mathematical description of the problem, assume all objectives need to be minimized (without loss of generality). The vector $U = [u_1, u_2, ..., u_k] \in \Re^k$ is dominant to vector $V = [v_1, v_2, ..., v_k] \in \Re^k$ (denoted by $U \prec V$) if and only if:

$$\forall i \in \{1, 2, ..., k\}, u_i \leq v_i \land \exists j \in \{1, 2, ..., k\} : u_j \leq v_j \tag{34}$$

In other words, there exists at least one $u_j$ smaller than $v_j$, such that the remaining $u$ values are smaller or equal to $v$ values.

The point $X^* \in \Omega$ ($\Omega$ is a feasible region in $\Re^n$) is considered an optimal Pareto (at least) with respect to $X \in \Omega$ if and only if $F(X^*) < F(X)$. Alternatively, it can be conveniently reconsidered that:

$$\forall X \in \Omega - \{X^*\} f_i(X^*) \leq f_i(X) \land \exists j \in \{1, 2, ..., k\} : f_j(X^*) < f_j(X) \tag{35}$$

In other words, the solution $X^*$ is said to be Pareto-optimal (at least) if no other solution is found to dominate $X^*$ based on the definition of Pareto superiority. For a given multi-objective optimization problem, the Pareto set $P^*$ is defined as a set in the decision variable space containing all optimal Pareto vectors $P^* = \{X \in \Omega | \nexists X' \in \Omega : F(X') < F(X)\}$. In other words, no other $X'$ exists to be placed in $\Omega$ as a vector of variable decisions dominating every $X \in P^*$.

A Pareto front $PF^*$ is a set of vector of objective functions obtained using the decision variables in the Pareto set, that is $PF^* = \{F(X) = (f_1(x), f_2(x), ..., f_k(x)) : X \in P^*\}$. In other words, the Pareto front $PF^*$ is a set of vectors of objective functions mapped from $P^*$.

In addition to using a population-based search approach [27], evolutionary algorithms are very suitable for solving multi-objective optimization problems.

This study uses the modified NSGA-II to optimize VAWTs. This modified NSGA-II is similar to the standard NSGA-II, with the exception that the -eliminate diversity method is replaced by the crowding distance division method in the modified NSGA-II. Using this approach, all clones and/or individuals with the same unique values will be removed from the current population. The elimination threshold in the present study is set at 0.001; hence, all members of the Pareto front within a certain distance of the individual are eliminated. Similarly, in order to prevent different individuals from being eliminated in the design variable space when they have similar values in the space of objective functions, similar values must be applied both within the objective and design variable spaces. A random selection of individuals from the population is then used to replace the eliminated individuals. As a result of this method, a more effective investigation of the search space is possible [40,41].

The flowchart shown in Fig. (10) illustrates the process of establishing an artificial neural network and optimizing the wind turbine. Following the formulation of the problem, the design of the experiments, and the CFD simulations, the output data is used to construct the ANN structure. Datasets for training and testing are separated from the sample data. Using the least-squares method, the coefficients of the partial models are estimated. Based on the test set, the external criteria for each partial model are calculated, and the partial models with lower criterion values are used as input for the next layer or the final polynomials.

During the first stage of the optimization code, an initial population of N individuals is generated, and the objective functions are then assessed using polynomials. A non-dominant approach is used to classify the population using the ε-eliminate diversity approach. The offspring population is created using a genetic algorithm that includes selection, cross-over, and mutation operators. In order to form a new population, the objective functions are re-evaluated, and the offspring and current population are combined. N individuals are selected from the new population after it has been sorted again. Based on the comparison of different objectives, an individual may be considered non-dominant if they are superior to other individuals. As the cost function of individuals is assigned, individuals in Pareto are entirely non-dominant, as individuals in the second front are only dominated by individuals in the first front. Priority is given to individuals from the first Pareto. In the event that the designed population size is not reached, the individuals from the second Pareto will be considered until the number of individuals has been reached in order to begin the next iteration.

Pareto optimal solutions are reported when their satisfaction criterion is not met. In such a case, the GA operators are re-applied, and the process is repeated to produce the next generations in order to obtain the Pareto optimal solution. The modified NSGA-II based on non-dominated sorting was applied for multi-objective optimization of the turbine performance, taking into account the above-mentioned geometric and evaluation parameters. A multi-objective optimization of the performance was then performed using polynomial ANNs. Based on the design variables (twist angle, aspect ratio, and overlap ratio), three objective functions are optimized: torque coefficient, rotational speed, and power coefficient. A modified NSGA-II is used to perform Pareto multi-objective evolutionary optimization. In order to evaluate the problem, the following formulation is used:

$$\text{Objective Functions} \begin{cases} f_1 = \text{Torque coefficient} & \text{maximize} \\ f_2 = \text{Rotational speed} & \text{maximize} \\ f_3 = \text{Power coefficient} & \text{maximize} \end{cases} \quad (36)$$

$$\text{Design Variables} \begin{cases} 0\ deg \leq Twist\ angle \leq 60\ deg \\ 0.8 \leq Aspect\ ratio \leq 1.2 \\ 0 \leq Overlab\ ratio \leq 0.24 \end{cases}$$

Note that instead of maximizing the torque coefficient, rotational speed, and power coefficient, the inverse functions, namely $1/C_T$, $1/$, and $1/C_P$, are minimized. For the three-objective optimization problem in 1000 generations, a population size of 200 is assumed with cross-over and mutation probabilities of 0.75 and 0.075, respectively. It is possible to plot individuals in different objective function planes as a result of solving this three-objective optimization problem. Using the same design variables, two-objective and multi-objective optimization problems are compared.

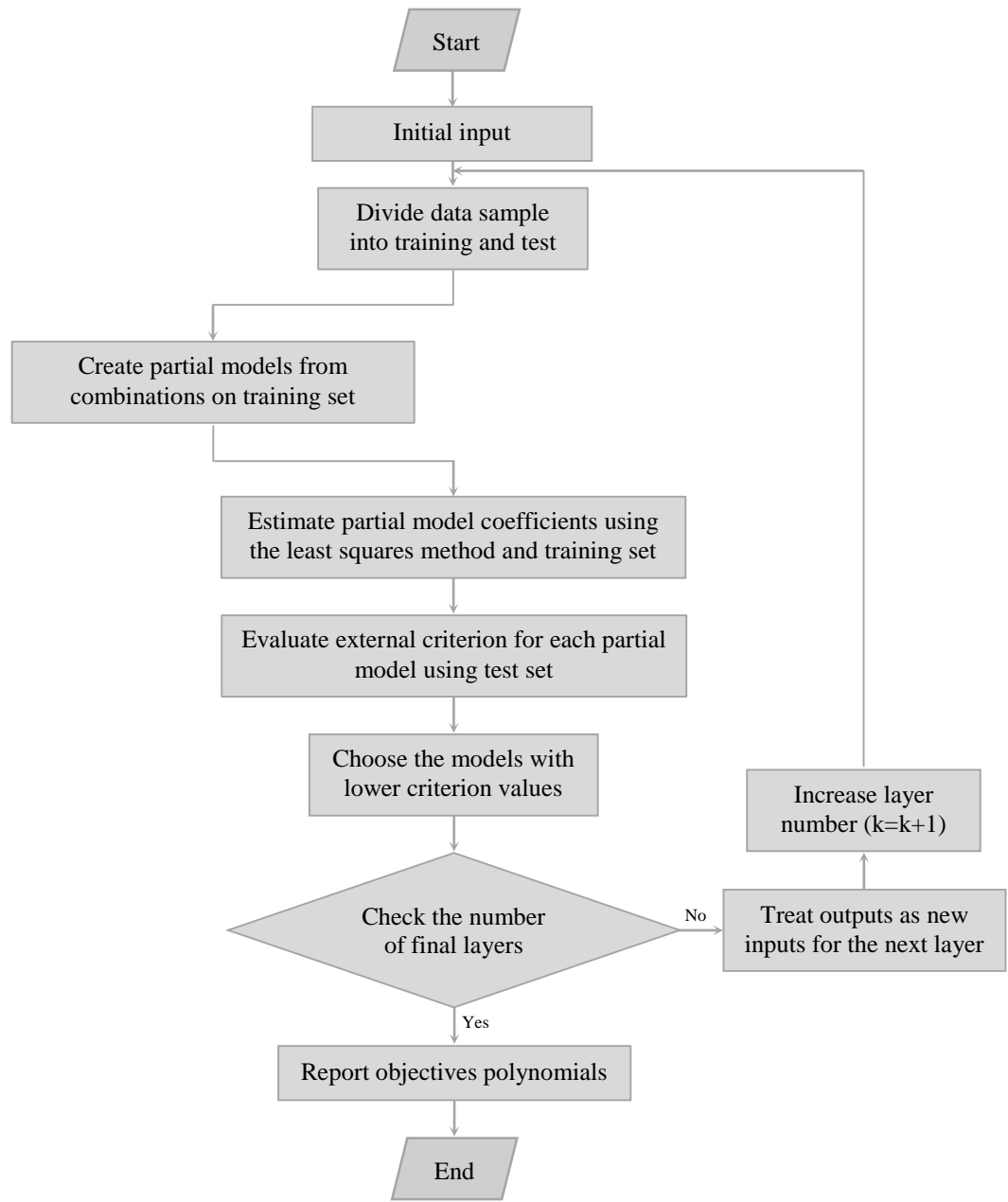

(a)

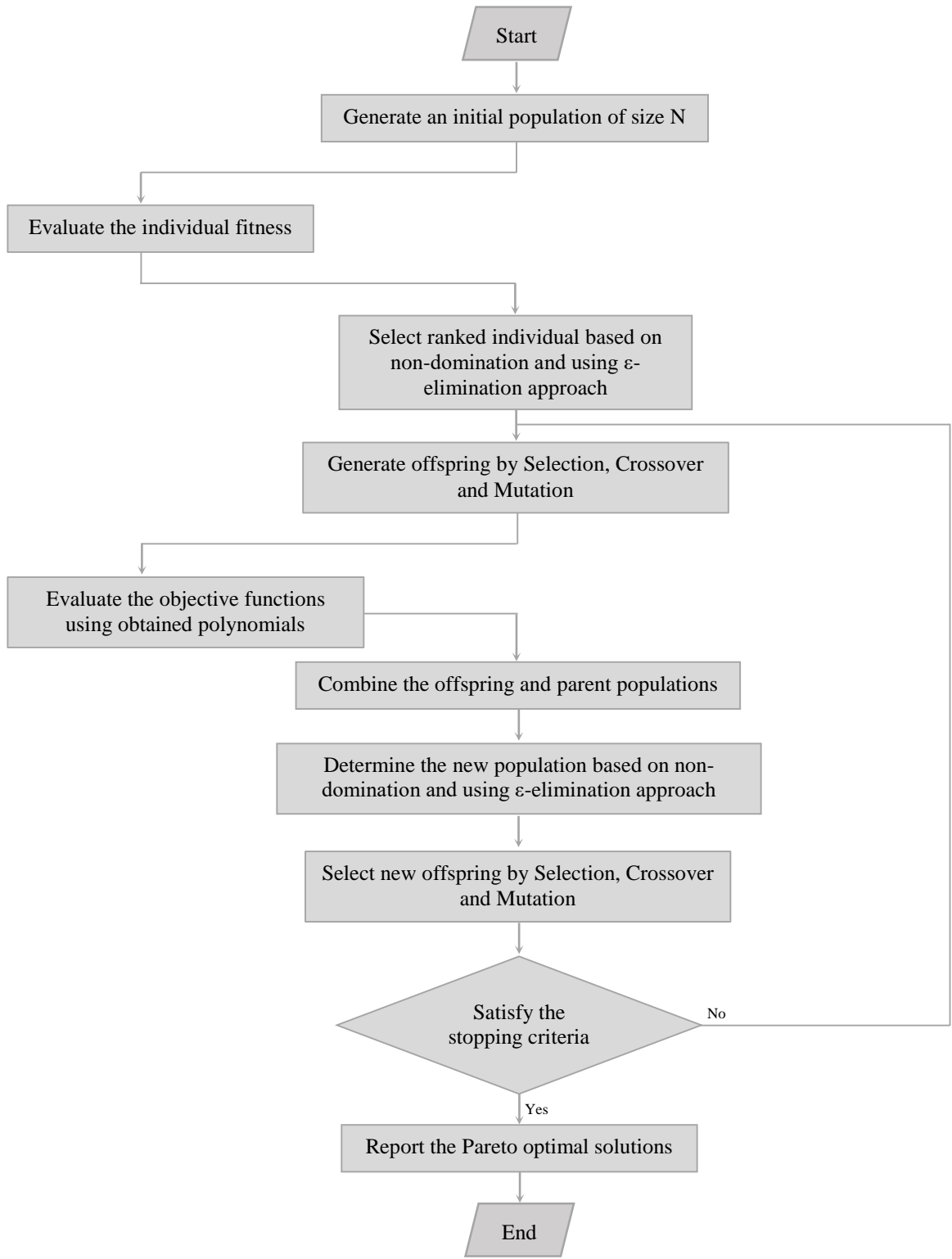

(b)

**Fig. 10** The flowchart of the wind turbine (a) artificial neural network and (b) optimization procedure.

## 7. TOPSIS

It is generally accepted that TOPSIS is a multi-criterion decision-making analysis (MCA) method that ranks the alternatives of the obtained Pareto solutions. Based on the target weight determined by the designer, the main objective of this method is to find the best compromise solution. The Multiple Attribute Decision-Making (MADM) method involves analyzing a given problem using an n×m matrix with m indices and n attributes. Each problem can be viewed as a geometric system consisting of m points in an n-dimensional space. As a result of this technique, each selected factor should have the smallest distance from the dominant factor (the positive ideal factor) and the highest distance from the least important factor (the negative ideal factor). The most significant advantages of this method can be summarized as follows: simultaneous combination (integration) of quantitative and qualitative criteria in evaluating the results, simplicity, appropriate speed, and consideration of the distance between the best and worst results from the optimal solution simultaneously.

An alternative solution should have the shortest distance from the positive ideal solution ($A^+$, i.e. the highest performance in each index) and also the farthest distance from the negative ideal solution ($A^-$, i.e. the lowest performance in each index). Moreover, the standard matrix $(x_{ij})_{m \times n}$ should be formed to select the best solution comparison where $x_{ij}$ denotes the $j^{th}$ target value of $i^{th}$ index. The resulting matrix should be then normalized using Eq. (37) to form the matrix $R$.

$$R = (r_{ij})_{m \times n}$$

$$r_{ij} = \frac{x_{ij}}{\sqrt{\sum_{i=1}^{m} x_{ij}^2}}, i = 1, 2, ..., m, j = 1, 2, ..., n \tag{37}$$

The weight of each index is obtained from the following equation by forming the weighted diagonal matrix:

$$T = (t_{ij})_{m \times n} = (w_j r_{ij})_{m \times n}, i = 1, 2, ..., m \qquad (38)$$

where $w_j$ is the weight assigned by the designer and, therefore $\sum_{j=1}^{n} w_j = 1$.

The positive and negative ideal solutions ($A^+$ and $A^-$) are obtained as follows:

$$A^+ = \{\langle \max(t_{ij} | i = 1, 2, ..., m) | j \in J_-\rangle, \langle \min(t_{ij} | i = 1, 2, ..., m) | j \in J_+\rangle\} \equiv \{t_{wj} | j = 1, 2, ..., n\}$$
$$A^- = \{\langle \min(t_{ij} | i = 1, 2, ..., m) | j \in J_-\rangle, \langle \max(t_{ij} | i = 1, 2, ..., m) | j \in J_+\rangle\} \equiv \{t_{bj} | j = 1, 2, ..., n\} \qquad (39)$$

where $J_+ = \{j = 1, 2, ..., n | j\}$ and $J_- = \{j = 1, 2, ..., n | j\}$ are respectively associated with positive and negative criteria.

To calculate the distance of $i^{\text{th}}$ attribute, the ratio of $d_{ib}$ with respect to the positive and negative ideal solutions ($d_{iw}$) is obtained as follows:

$$d_{ib} = \sqrt{\sum_{j=1}^{n} (t_{ij} - t_{bj})^2}, i = 1, 2, ..., m$$
$$d_{iw} = \sqrt{\sum_{j=1}^{n} (t_{ij} - t_{wj})^2}, i = 1, 2, ..., m \qquad (40)$$

Eventually, the relative proximity to the ideal solution is determined based on $s_{iw} = d_{iw} / (d_{iw} + d_{ib}), 0 \leq s_{iw} \leq 1, i = 1, 2, ..., m$ oscillating between 0 and 1.

## 8. Results and discussion

The dominated optimal design points obtained from the three- and two-objective optimizations overlap, as shown in Fig. (11a). Design points are also shown in Figs. (11b) and (11c) on other planes. Since front points in Pareto solutions are regarded as the best points, their corresponding design variables are regarded as the most desirable. The two-objective values corresponding to another set of design variables would rank lower than the Pareto front if another set of design variables were selected. The Pareto set provides the best combination of all three objectives when the design variables are selected according to the Pareto set. In each plane, all the Pareto front points dominate each other, but all are superior to the others.

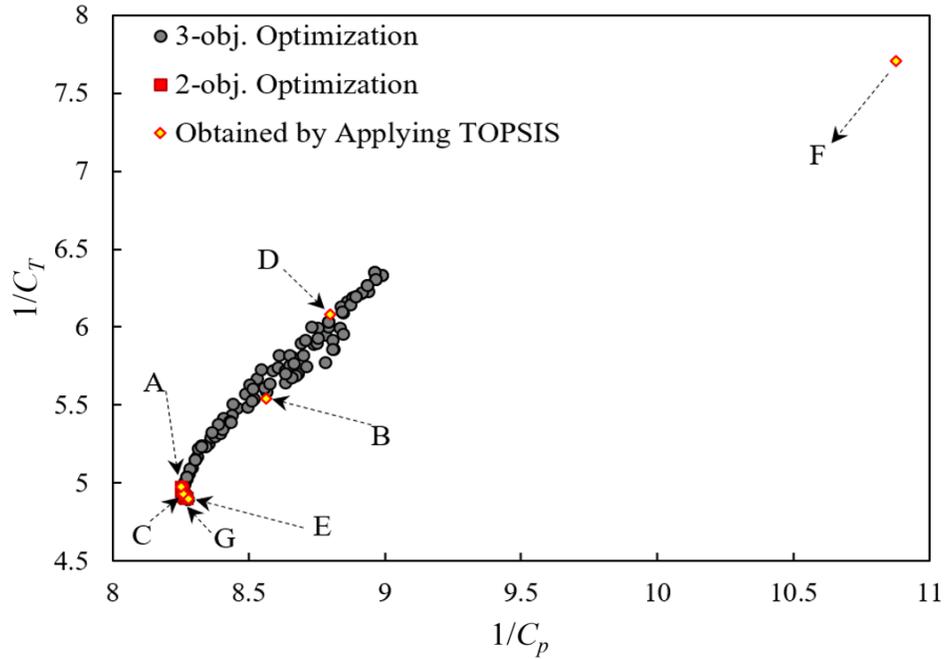

(a)

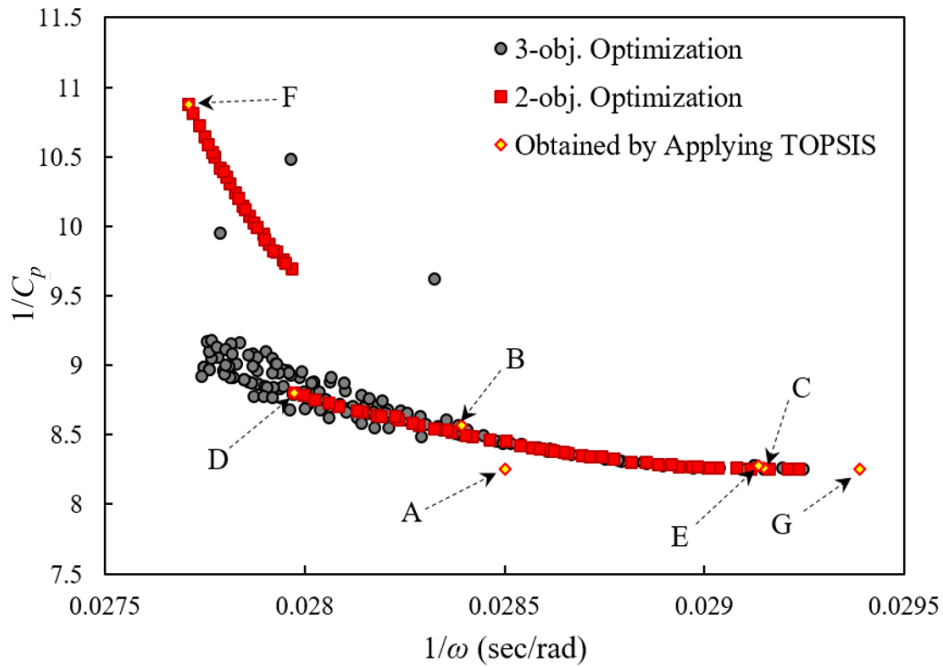

(b)

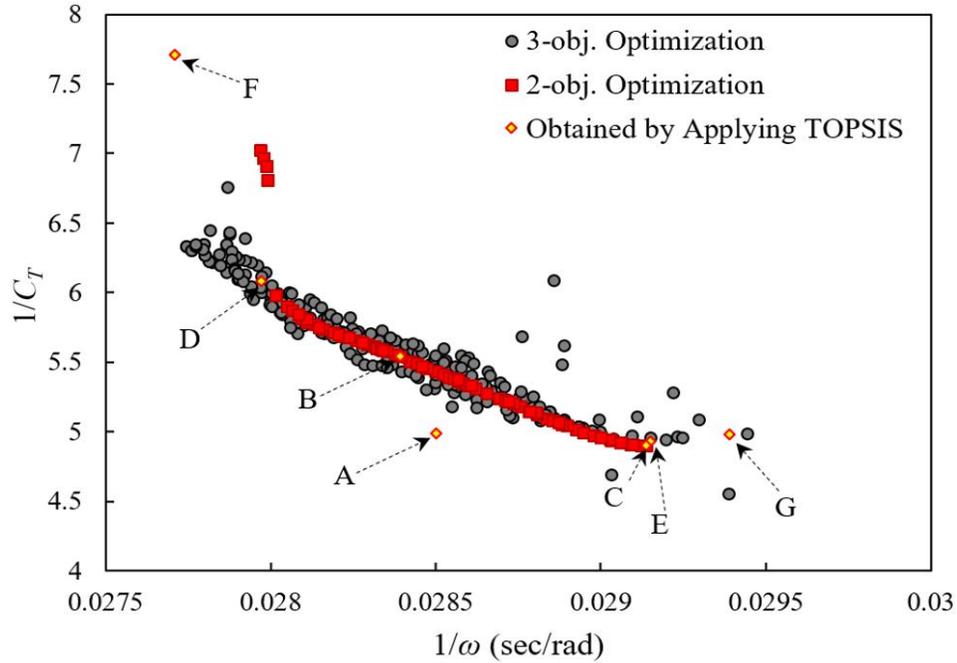

(c)

**Fig. 11** (a) The non-dominated optimum design points in the plane of $1/C_T$ and $1/C_p$ (b) plane of $1/C_p$ and $1/\omega$, (c) and plane of $1/C_T$ and $1/\omega$.

As shown in Fig. (11), the results of three-objective optimization in each plane include the results obtained from a two-objective problem, providing designers with more choices. Furthermore, the results of the two-objective optimization are located at the boundary of the three-objective problem, indicating the validity of the results.

According to Fig. (11), the values of 1/CT obtained using the two-objective problem are the same as those obtained using the modified NSGA-II. As a result, it is evident that selecting $1/C_p$ as an objective function gives the same results as choosing $1/C_T$; therefore, they can be used interchangeably. 1/rotational speed is the result of 1/pressure coefficient and 1/torque coefficient drops. As the 1/pressure coefficient increases, the 1/torque coefficient increases, but pressure has a much lower contribution to rotational speed than torque. Therefore, in the wind turbine under consideration, torque dominates rotational speed.

For the purpose of determining the optimal points at which all of the objective functions are jeopardized, the TOPSIS method was employed. Each diagram shows the optimal design points. A list of objective function values can be found in Table 8. The TOPSIS method was used to investigate the design variables of the obtained points. In this case, Point A is obtained by applying the TOPSIS algorithm to the results of a three-objective optimization problem. In the selected objective functions, the Pareto exponents obtained from two different objective functions are clearly visible. The use of TOPSIS for these Pareto fronts results in deviations from the optimal design points (B to D), compromising both objective functions.

**Table 8** The values of the design variables and objective functions of the optimum points specified using TOPSIS.

|  | Point | Objective functions | $\varphi$ (deg) | $\alpha$ | $\delta$ | $1/C_T$ | $1/\omega$ (sec/Rad) | $1/C_p$ |
|---|---|---|---|---|---|---|---|---|
| Three-objective | A | $1/C_t$-$1/\omega$-$1/C_p$ | 4.57 | 0.835 | 0.136 | 4.98 | 0.0285 | 8.25 |
| Two-objective | B | $1/C_t$-$1/\omega$ | 9.47 | 0.949 | 0.134 | 5.54 | 0.0283 | 8.56 |
|  | C | $1/C_t$-$1/C_p$ | 32.98 | 0.964 | 0.138 | 4.92 | 0.0291 | 8.25 |
|  | D | $1/\omega$-$1/C_p$ | 4.48 | 0.861 | 0.142 | 6.08 | 0.0279 | 8.79 |
| single-objective | E | $1/C_t$ | 36.56 | 0.972 | 0.12 | 4.9 | 0.0291 | 8.28 |
|  | F | $1/\omega$ | 3.83 | 1.181 | 0.155 | 7.7 | 0.0277 | 10.87 |
|  | G | $1/C_p$ | 31.4 | 0.963 | 0.154 | 4.97 | 0.0293 | 8.25 |

Torque and power coefficients are improved by 13.83% and 5.30%, respectively, when considering both the torque and power coefficients as independent objective functions (Table 9).

**Table 9** The baseline values of objective functions.

| Parameters | $1/C_T$ | $1/\omega$ (sec/Rad) | $1/C_p$ |
|---|---|---|---|
| Values | 5.71 | 0.028 | 8.72 |

As can also be seen in Fig. (11), single-objective optimization results (Points E to G) indicate the most optimal values for each objective. By considering 1/CT as a single-objective function, the lowest value of 1/CT (4.9) can be obtained. 1/(rotational speed) and 1/(power coefficient) are minimized to obtain the maximum rotational speed (36.1) and power coefficient (0.121). Furthermore, the closest point to the ideal point method can be used to determine the optimal trade-off point of the optimal design. Using this method, the objective function values for all non-dominating points are mapped from 0 to 1, and the distance between each optimal point and the ideal point is calculated. The optimal design point 39 is determined by the minimum distance to the ideal point. Using the design point (Table 10) obtained from this method, the torque coefficient, rotational speed, and power coefficient are improved by 13.74%, 0.071%, and 5.32%, respectively.

**Table 10** The values of the design variables and objective functions of the optimum point specified using nearest to the ideal point method.

| Design variables | | | Objective functions | | |
| --- | --- | --- | --- | --- | --- |
| $\varphi$ (deg) | $\alpha$ | $\delta$ | $1/C_T$ | $1/\omega$ (sec/Rad) | $1/C_p$ |
| 6.81 | 0.903 | 0.129 | 4.925 | 0.0278 | 8.256 |

The optimization process significantly improves the power coefficient regardless of the method used for selecting the trade-off points of the optimal design, as shown in Table 10. Thus, reducing undesirable forces can help reduce depreciation and related costs by determining the optimal operating point of the VAWT. On the basis of the three-objective optimization method, the torque coefficient was improved by 14.18% and 13.74%, respectively. As a result of all the optimized trade-off points obtained, the twist angle has been reduced as compared to its initial value.

Figure (12) compares the pressure and fluid velocity of the original and optimized turbines. In Fig. (12 a), the original turbine has a larger high-pressure area than the optimized turbine, but the modified turbine has a higher pressure gradient across the driving blade, which results in higher torque.

Figure (12 b) illustrates backward flows toward retarded blades in both turbines, with the backflow being stronger for the optimized turbine, resulting in an improved torque. Additionally, the overlap ratio enhances the fluid's ability to enter and exit the turbine more smoothly, especially in the optimized turbine. Therefore, the optimized turbine will experience fewer losses and blockages, which will improve the performance of the turbine as a whole.

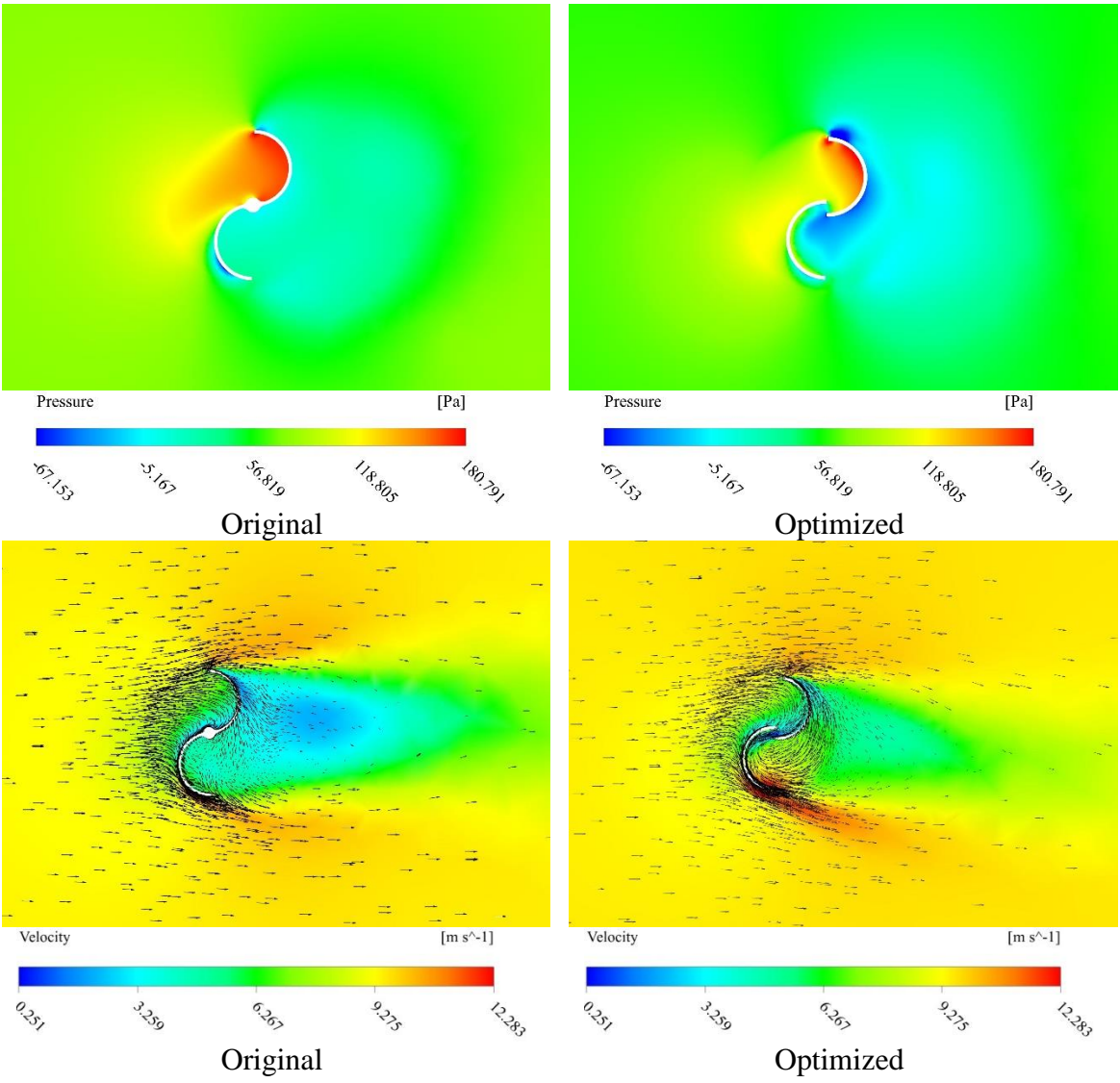

**Fig. 12** Comparison between the original and optimized turbines. a) Pressure distributions around turbines; b) The fluid velocity vectors.

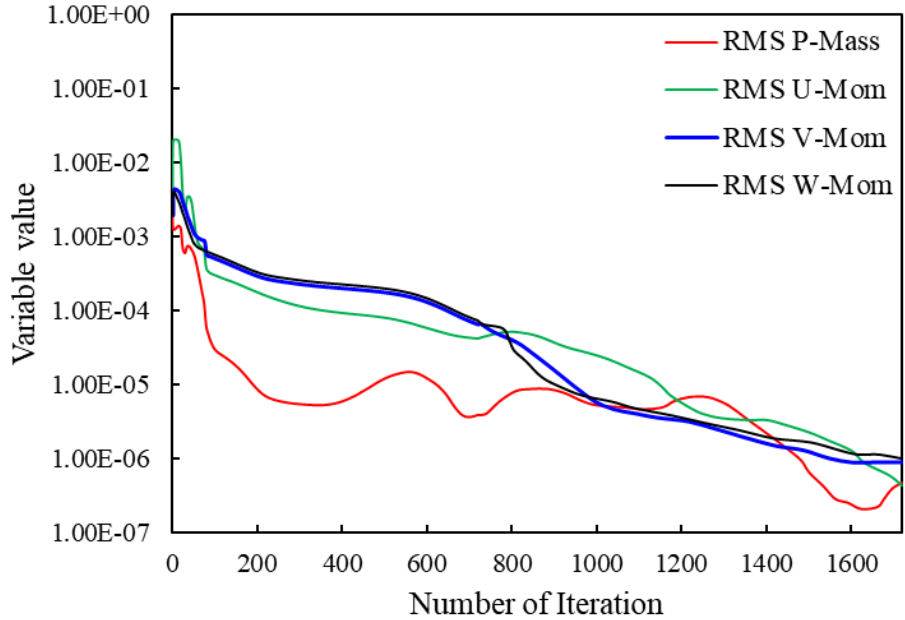

Fig 13. Convergence history of results

To provide a comprehensive view of the relationship between objective functions and design variables, Figs. (14) to (16) illustrate optimal variations of the objective functions with respect to the design variables. The values of all design variables are distributed across their permissible domains.

According to figures (14-16), the blade twist angle varies almost linearly; however, the aspect ratio and overlap ratio are not constant. This paper presents a multi-objective Pareto optimization procedure that discovers relationships that are indefeasible between wind turbine optimal design variables. By using this optimization method, designers can select the desired design parameters according to their requirements.

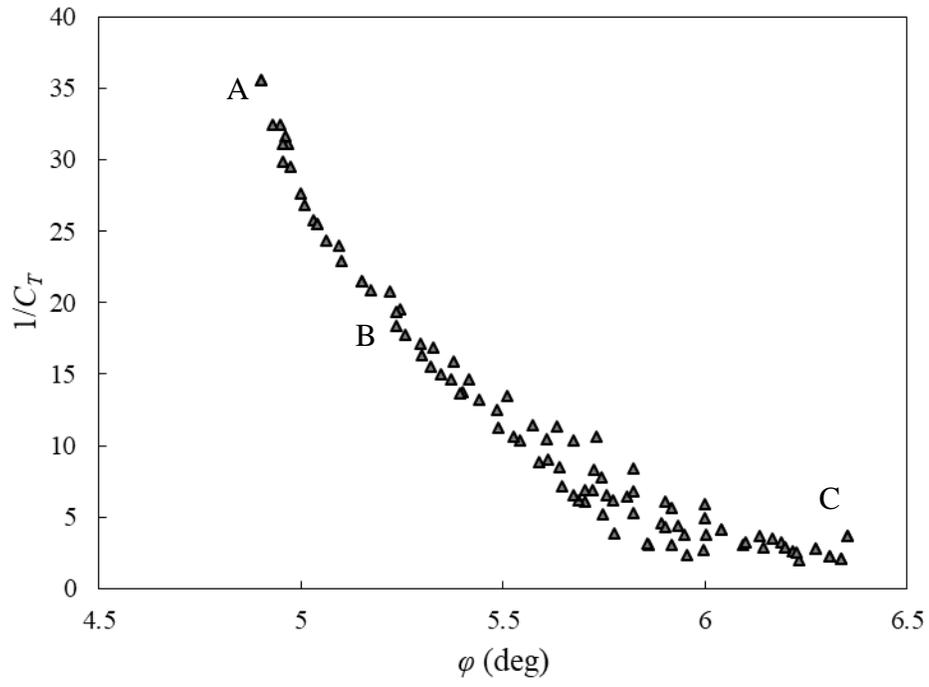

(a)

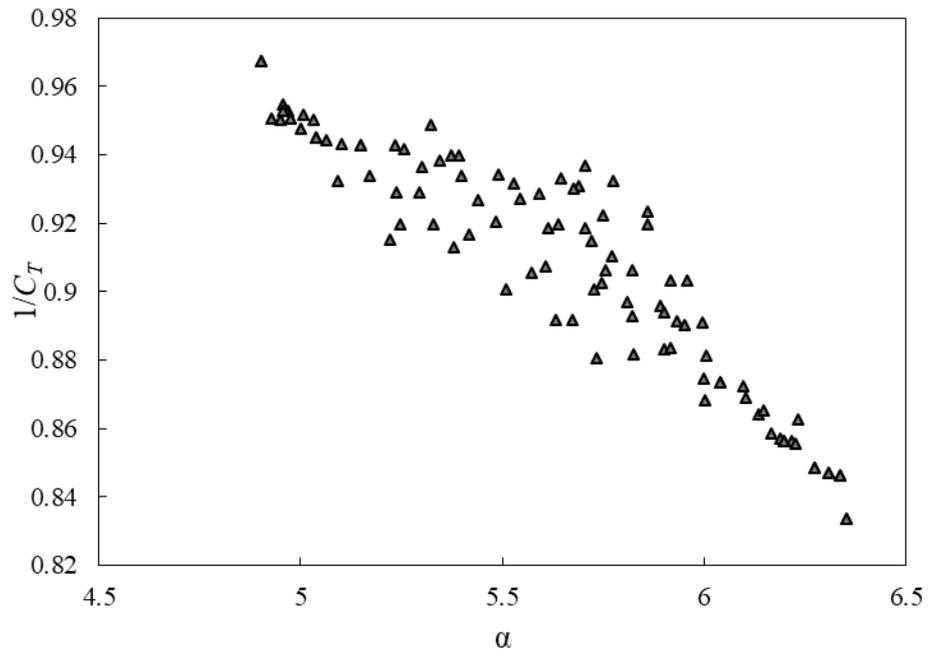

(b)

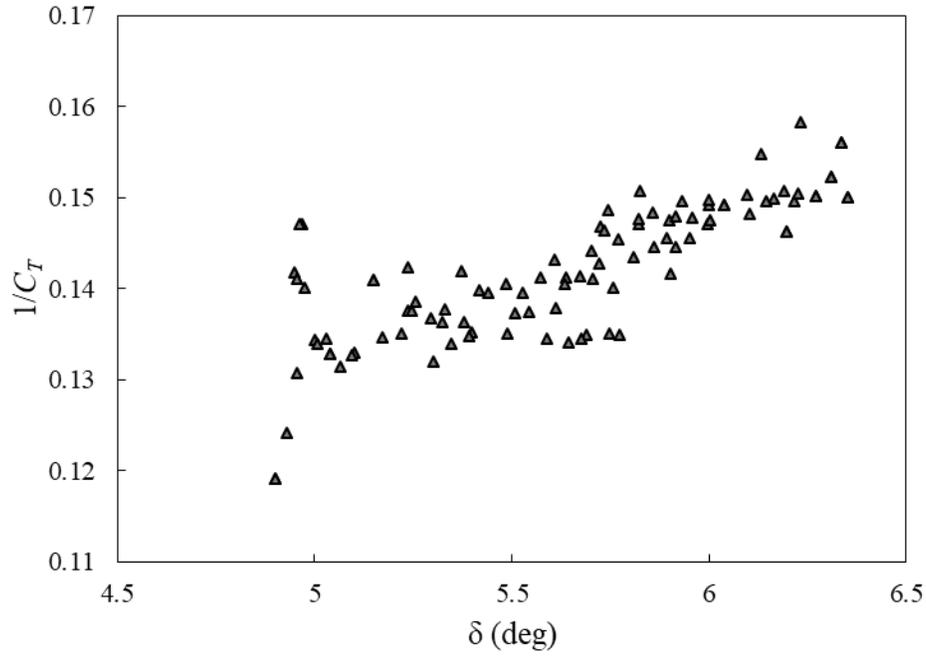

(c)

**Fig. 14** Variations of the $1/C_T$ versus (a) the blade twist angle, (b) the aspect ratio and (c) the overlap ratio.

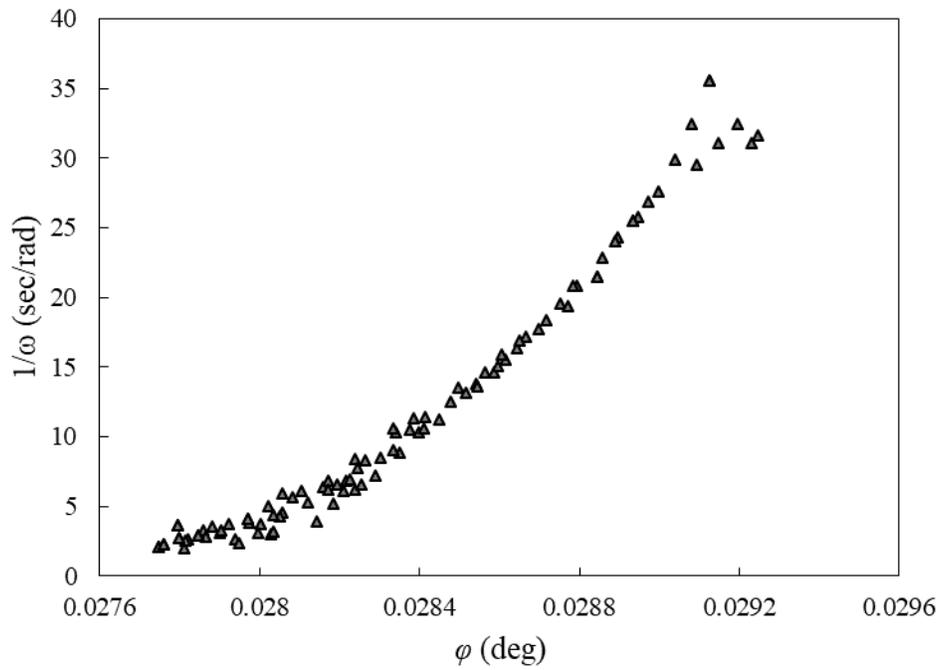

(a)

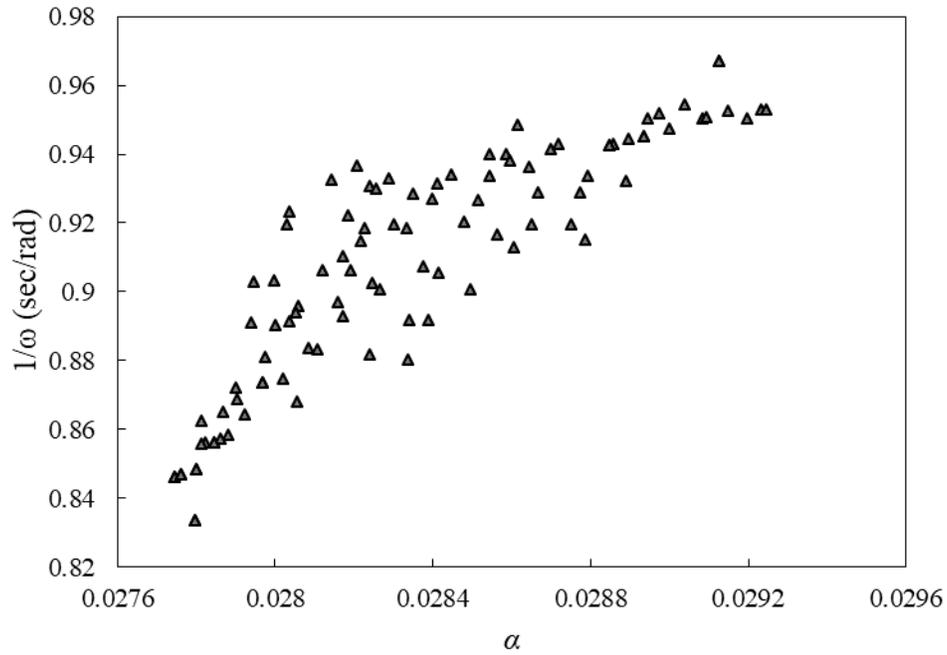

(b)

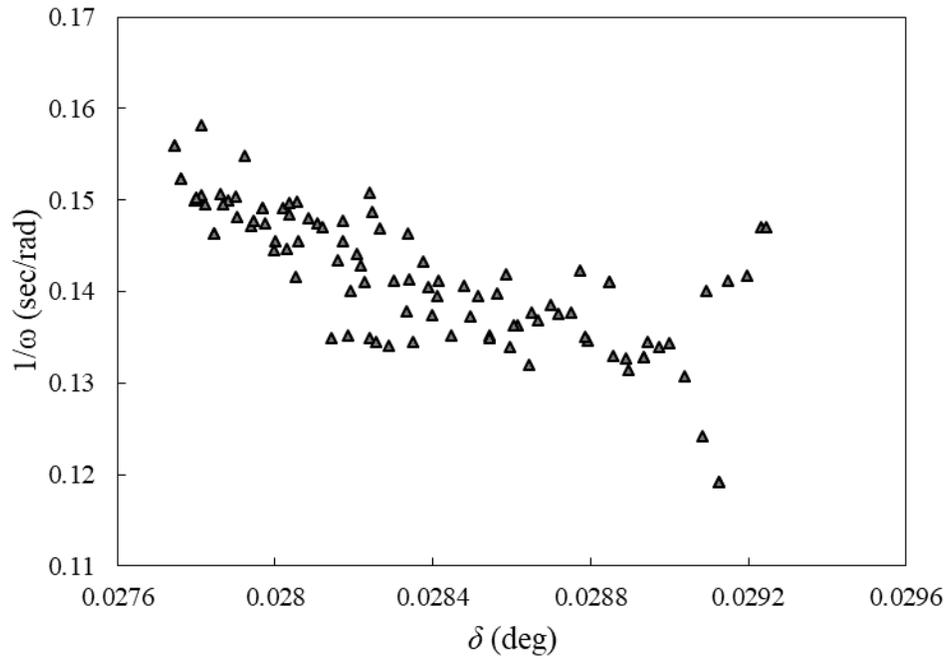

(c)

**Fig. 15** Variations of the $1/\omega$ versus (a) the blade twist angle, (b) the aspect ratio and (c) the overlap ratio.

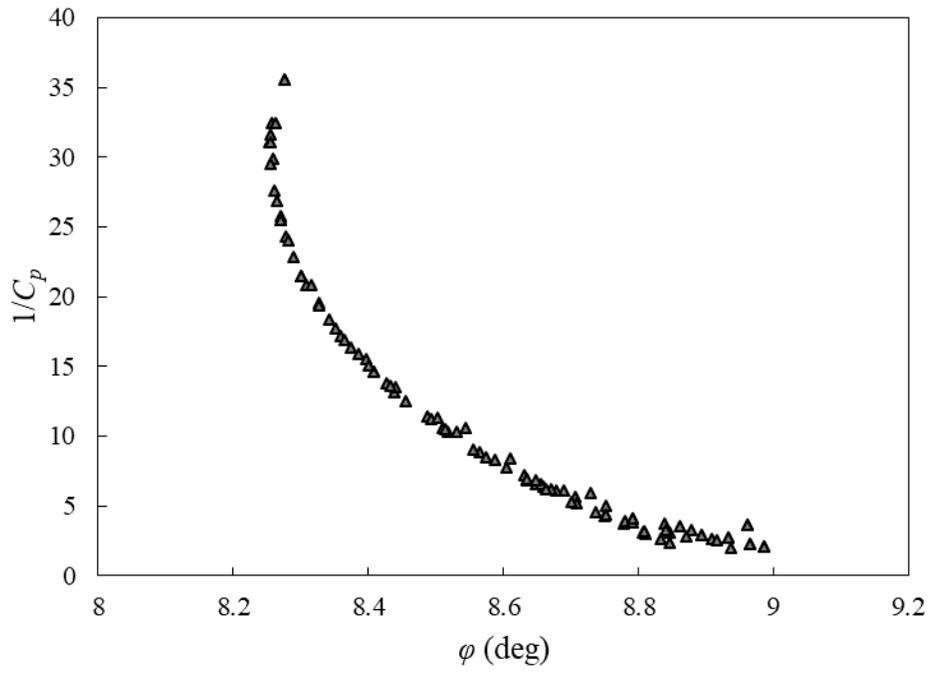

(a)

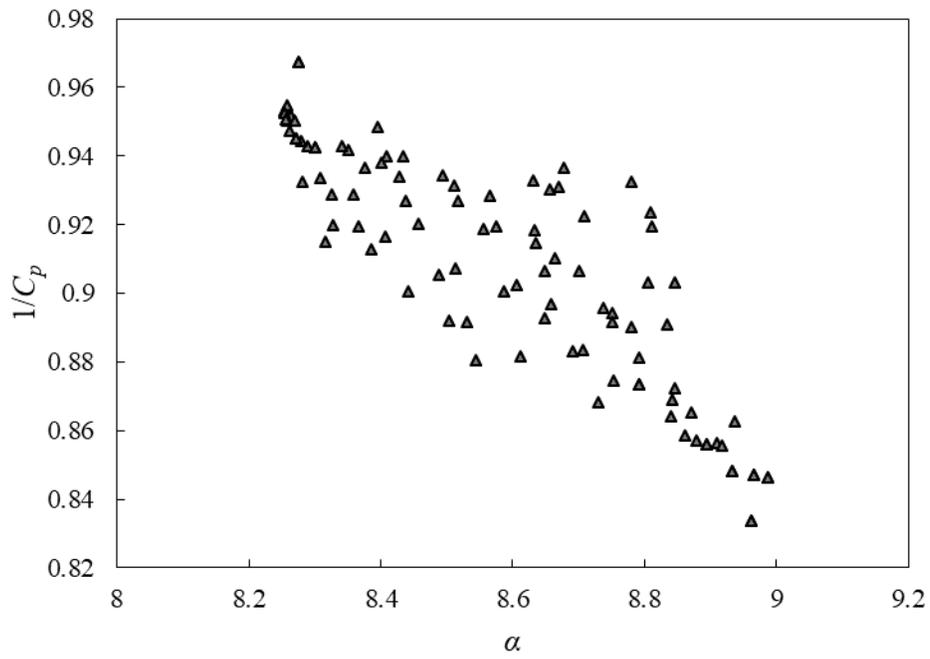

(b)

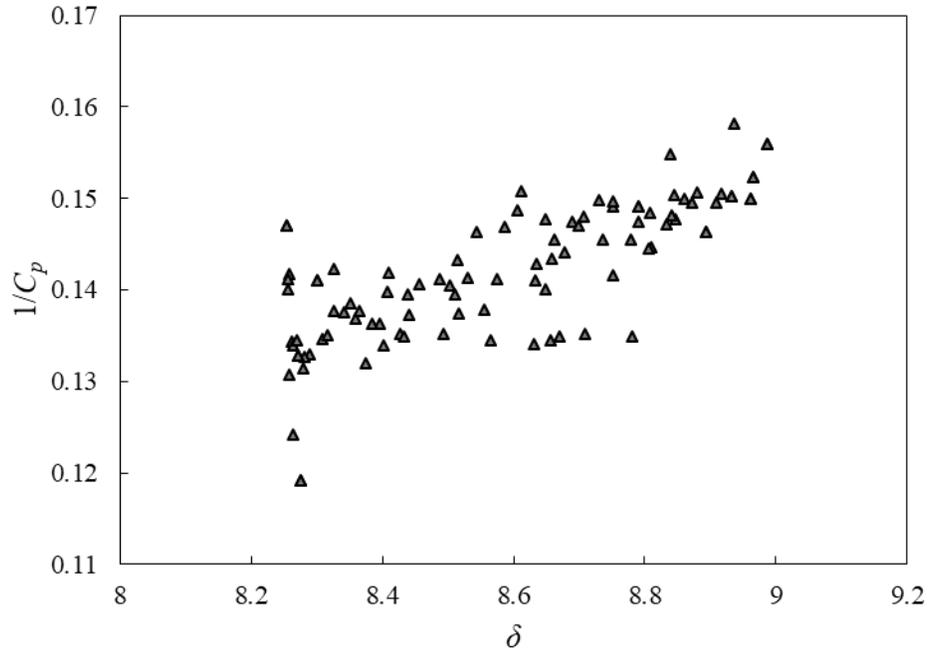

(c)

**Fig. 16** Variations of the $1/C_p$ versus (a) the blade twist angle, (b) the aspect ratio and (c) the overlap ratio.

In each region, one design variable changes for all figures, while the other variables remain almost constant. Furthermore, in the middle region, all objectives are competing, so this region is expected to contain the final design point (Point A in Table 8).

Due to the absence of these points in the training and test sets, TOPSIS' optimal points are reevaluated by CFD. In Table 11, the CFD results are compared with those obtained from the GMDH. According to the results of the GMDH, there is a good agreement between the CFD and GMDH results, indicating that the GMDH is an accurate model.

**Table 11** Comparison between CFD and GMDH predicted values of the optimum points.

| Point | 1/$C_T$ | | | 1/$\omega$ (sec/Rad) | | | 1/$C_p$ | | |
|---|---|---|---|---|---|---|---|---|---|
| | GMDH | CFD | Absolute error (%) | GMDH | CFD | Absolute error (%) | GMDH | CFD | Absolute error (%) |
| A | 4.98 | 4.59 | 7.83 | 0.0285 | 0.0287 | 0.7 | 8.25 | 7.86 | 4.72 |
| B | 5.54 | 5.49 | 0.9 | 0.0283 | 0.028 | 1.06 | 8.56 | 8.31 | 2.92 |
| C | 4.92 | 4.93 | 0.2 | 0.0291 | 0.0286 | 1.71 | 8.25 | 8.42 | 2.06 |
| D | 6.08 | 6.17 | 1.48 | 0.0279 | 0.0291 | 4.3 | 8.79 | 9.12 | 3.75 |
| E | 4.9 | 4.52 | 7.75 | 0.0291 | 0.02796 | 3.91 | 8.28 | 9.02 | 8.93 |
| F | 7.7 | 7.26 | 5.71 | 0.0277 | 0.0283 | 2.16 | 10.87 | 9.93 | 8.64 |
| G | 4.97 | 4.98 | 0.2 | 0.0293 | 0.02761 | 5.76 | 8.25 | 8.24 | 0.12 |

With the use of a multi-objective optimization method, it would be impossible to consider all the competition criteria in the design of the VAWT and also discover the relationship between the optimal design variables.

## 9. Conclusions

To evaluate the performance of a twisted Savonius wind turbine, both numerical simulations and experimental tests were conducted. The average deviation between numerical and experimental results, at a pressure difference of 16.5 percent, was approximately 1.65% at a wind speed of 7 m/s, indicating that the numerical method employed in this study was quite effective. In addition, a multi-objective optimization technique was used to optimize the performance of the turbine. To increase the power coefficient, torque coefficient, and rotational speed, twist angle, aspect ratio, and overlap ratio were considered to be design variables. Input-output data were used to model the objective functions using the GMDH neural network. The polynomial models derived from the evolutionary Pareto-based optimization approach (the modified NSGA-II) were used to plot Pareto fronts, and TOPSIS was used to determine the optimal commercial points. Comparison of three- and two-objective optimization data indicates that two-objective optimization data fall within the boundaries of a three-objective problem. Multi-objective optimization improved the torque coefficient, rotational speed, and power coefficient by 13.74%, 0.071%, and 5.32%, respectively. Several important characteristics of the objective functions were discovered during the multi-objective optimization of VAWT. Turbine design can benefit greatly from the combination of TOPSIS and NSGA-II. Based on our research, we believe there are four important points to be made:

1. With an overlap ratio of 0.16, an aspect ratio of 1, and a twist angle of 45°, the highest power coefficient of 0.136 was obtained. As a result, air is able to escape through the blades, preventing the formation of the drag force that resists the movement.
2. The power coefficient initially decreased as the overlap ratio increased from 0 to 0.16, but then increased and reached its maximum value as the overlap ratio increased from 0 to 0.16.
3. In response to an increase in the overlap ratio, the power coefficient again experienced a decreasing trend with a higher gradient.
4. We find that the highest power coefficient is obtained with an aspect ratio of 1 and a twist angle of 45 degrees.

**Conflict of Interest:** The authors declare that they have no conflict of interest.

**Funding:** There is no funding source.

**Ethical approval:** This article does not contain any studies with human participants or animals performed by any of the authors.

**Nomenclature**

| | | | |
|---|---|---|---|
| $k$ | Turbulent kinetic energy ($m^2/s^2$) | $F_1$ | Blending function composition |
| $U_j$ | Mean velocity (m/s) | $y$ | Distance to the nearest wall |
| $P$ | Pressure (Pa) | $\tau$ | Stress tensor |
| $u'$ | Fluctuating velocity (m/s) | $\lambda$ | Tip speed ratio |
| $\rho$ | Density (kg/$m^3$) | $\delta$ | Overlap ratio |
| $d$ | Diameter of each blade | $\omega$ | Rotational speed |
| $\Omega$ | absolute value of vorticity | $\mu$ | Fluid viscosity |
| $P_w$ | Turbine power | $T$ | Produced torque |
| $P_a$ | Wind power | $T_w$ | Theoritical torque |
| $D$ | Turbine diameter | $\mu_t$ | Turbulent viscosity (Pa s) |
| $h$ | Turbine height | **Acronyms** | |
| $A_s$ | Vertical mirrored area | *CFD* | Computational fluid dynamics |
| $\alpha$ | Aspect ratio | *RANS* | Reynolds-Averaged Navier–Stokes |
| $\varphi$ | Twist angle | TSR | Tip speed ratio |
| $C_p$ | Power coefficient | ANN | Artificail neural network |
| $C_T$ | Torque coefficient | VAWT | Vertical-axis wind turbine |
| $P_a$ | Wind power | GA | Genetic algorithm |
| $\delta_{ij}$ | Kronecker delta | | |

**Appendix 1**

**Table 5** Samples of numerical results using CFD.

|  | Input data | | | | Output data | | |
|---|---|---|---|---|---|---|---|
|  | $\varphi$ | $\alpha$ | $\delta$ | | $C_p$ | $C_T$ | $\omega$ |
| 1 | 0 | 0.8 | 0 | | 0.10735 | 0.142833 | 37.98526 |
| 2 | 15 | 0.8 | 0 | | 0.11875 | 0.161161 | 37.24395 |
| 3 | 30 | 0.8 | 0 | | 0.119894 | 0.165901 | 36.176 |
| 4 | 45 | 0.8 | 0 | | 0.12735 | 0.174851 | 35.6972 |
| 5 | 60 | 0.8 | 0 | | 0.10807 | 0.145388 | 35.0056 |
| 6 | 0 | 1 | 0 | | 0.16414 | 0.154274 | 37.3464 |
| 7 | 15 | 1 | 0 | | 0.11678 | 0.162188 | 36.71072 |
| 8 | 30 | 1 | 0 | | 0.12065 | 0.173663 | 35.112 |
| 9 | 45 | 1 | 0 | | 0.12902 | 0.192382 | 33.9416 |
| 10 | 60 | 1 | 0 | | 0.1083 | 0.164091 | 33.3564 |
| 11 | 0 | 1.2 | 0 | | 0.0969 | 0.129838 | 37.7188 |
| 12 | 15 | 1.2 | 0 | | 0.09975 | 0.134606 | 37.4528 |
| 13 | 30 | 1.2 | 0 | | 0.1064 | 0.1444 | 37.24 |
| 14 | 45 | 1.2 | 0 | | 0.11495 | 0.158035 | 36.7612 |
| 15 | 60 | 1.2 | 0 | | 0.1045 | 0.145993 | 36.176 |
| 16 | 0 | 0.8 | 0.08 | | 0.09944 | 0.128991 | 33.43155 |
| 17 | 15 | 0.8 | 0.08 | | 0.11 | 0.144954 | 32.90918 |
| 18 | 30 | 0.8 | 0.08 | | 0.11 | 0.148973 | 32.02116 |
| 19 | 45 | 0.8 | 0.08 | | 0.1144 | 0.158289 | 31.34208 |
| 20 | 60 | 0.8 | 0.08 | | 0.09328 | 0.131032 | 30.87195 |
| 21 | 0 | 1 | 0.08 | | 0.1056 | 0.139406 | 32.84995 |
| 22 | 15 | 1 | 0.08 | | 0.10912 | 0.144482 | 32.75247 |
| 23 | 30 | 1 | 0.08 | | 0.11176 | 0.154379 | 31.39432 |
| 24 | 45 | 1 | 0.08 | | 0.11968 | 0.156074 | 33.25414 |
| 25 | 60 | 1 | 0.08 | | 0.10032 | 0.146113 | 29.77498 |
| 26 | 0 | 1.2 | 0.08 | | 0.08976 | 0.119971 | 32.44595 |
| 27 | 15 | 1.2 | 0.08 | | 0.0924 | 0.120234 | 33.32708 |
| 28 | 30 | 1.2 | 0.08 | | 0.09856 | 0.129673 | 32.96142 |
| 29 | 45 | 1.2 | 0.08 | | 0.10648 | 0.142349 | 32.43905 |
| 30 | 60 | 1.2 | 0.08 | | 0.0968 | 0.130036 | 32.28234 |
| 31 | 0 | 0.8 | 0.16 | | 0.113 | 0.18138 | 34.888 |
| 32 | 15 | 0.8 | 0.16 | | 0.125 | 0.198413 | 35.28 |
| 33 | 30 | 0.8 | 0.16 | | 0.125 | 0.203915 | 34.328 |
| 34 | 45 | 0.8 | 0.16 | | 0.13 | 0.218121 | 33.376 |
| 35 | 60 | 0.8 | 0.16 | | 0.106 | 0.179357 | 33.096 |
| 36 | 0 | 1 | 0.16 | | 0.12 | 0.187999 | 35.7448 |
| 37 | 15 | 1 | 0.16 | | 0.124 | 0.197767 | 35.112 |
| 38 | 30 | 1 | 0.16 | | 0.127 | 0.211314 | 33.656 |
| 39 | 45 | 1 | 0.16 | | 0.136 | 0.254206 | 29.96 |
| 40 | 60 | 1 | 0.16 | | 0.114 | 0.183871 | 34.72 |
| 41 | 0 | 1.2 | 0.16 | | 0.102 | 0.159375 | 35.84 |
| 42 | 15 | 1.2 | 0.16 | | 0.105 | 0.164577 | 35.728 |
| 43 | 30 | 1.2 | 0.16 | | 0.112 | 0.177496 | 35.336 |
| 44 | 45 | 1.2 | 0.16 | | 0.121 | 0.194847 | 34.776 |
| 45 | 60 | 1.2 | 0.16 | | 0.11 | 0.177994 | 34.608 |
| 46 | 0 | 0.8 | 0.24 | | 0.09605 | 0.125295 | 31.01616 |
| 47 | 15 | 0.8 | 0.24 | | 0.10625 | 0.147521 | 29.14072 |
| 48 | 30 | 0.8 | 0.24 | | 0.10625 | 0.156733 | 27.42807 |
| 49 | 45 | 0.8 | 0.24 | | 0.1105 | 0.166534 | 26.8464 |
| 50 | 60 | 0.8 | 0.24 | | 0.0901 | 0.157732 | 23.1117 |
| 51 | 0 | 1 | 0.24 | | 0.102 | 0.120416 | 34.2721 |
| 52 | 15 | 1 | 0.24 | | 0.1054 | 0.14012 | 30.43449 |
| 53 | 30 | 1 | 0.24 | | 0.10795 | 0.16242 | 26.89114 |
| 54 | 45 | 1 | 0.24 | | 0.1156 | 0.202514 | 23.09552 |
| 55 | 60 | 1 | 0.24 | | 0.0969 | 0.143044 | 27.40808 |
| 56 | 0 | 1.2 | 0.24 | | 0.0867 | 0.11668 | 30.06416 |
| 57 | 15 | 1.2 | 0.24 | | 0.08925 | 0.126497 | 28.54667 |
| 58 | 30 | 1.2 | 0.24 | | 0.0952 | 0.129858 | 29.66146 |
| 59 | 45 | 1.2 | 0.24 | | 0.10285 | 0.14724 | 28.26202 |
| 60 | 60 | 1.2 | 0.24 | | 0.0935 | 0.130091 | 29.07979 |